\documentclass[12pt]{article}
\usepackage{amsmath,amssymb}
\usepackage{wrapfig}
\usepackage{epsfig}
\usepackage{fancybox}
\usepackage{bm}
\usepackage{mathrsfs}
\usepackage{amsthm}
\usepackage{enumerate}
\usepackage{comment}
\usepackage{fourier}
\usepackage{multirow}
\allowdisplaybreaks[4]
\newcommand{{\Slash}}{0\!\!\!\!\!\big/}
\usepackage[usenames]{color}
\usepackage{setspace}
\usepackage {arydshln}

\begin{document}

\title{Embedded structure in quantum theory, 
functional operator and multiverse}

\author{
Yoshiharu \textsc{Kawamura}\footnote{E-mail: haru@azusa.shinshu-u.ac.jp}\\
{\it Department of Physics, Shinshu University, }\\
{\it Matsumoto 390-8621, Japan}\\
}


\maketitle
\begin{abstract}
We explore a wider theoretical framework that has quantum field theory built-in,
taking the fact that quantum mechanics is reconstructed from quantum field theory as a hint.
We formulate a quantum theory with an embedded structure
by introducing functional operators, 
and we find that it could describe the level II multiverse.
Topics related to a beginning of the universe such as an inflation, the third quantization
and the landscape are discussed in our formulation.
\end{abstract}

\section{Introduction}

Quantum field theory (QFT) or quantum theory of fields offers an excellent framework
that explains a great variety of phenomena in our world
and describes curious features such as a creation and annihilation of particles, 
the dichotomy between a particle and a wave, and so on~\cite{D,H&P}.
In fact, QFT has been applied to various systems and been massively successful.
For example, particle physics at the electroweak scale is excellently explained
by the standard model~\cite{SM},
and the superconductivity becomes better understood
by BCS theory~\cite{BCS} in condensed matter physics.

In spite of such prominent features and triumphs of QFT,
it cannot be an ultimate framework of physics
because it suffers from intrinsic problems.
For instance, QFT has a divergence difficulty 
that theoretical values of physical quantities
diverge to infinity after radiative corrections are incorporated.
In particular, this problem becomes serious after the gravitational interaction
is introduced, because there appear infinities
which cannot be removed by the renomalization procedure.
Hence, QFT is currently understood as an effective theory
of quantum fields~\cite{W}.

Furthermore, we have several questions in mind.
\begin{enumerate}[Q1]
\item Why does QFT work as an effective theory of elementary particles extremely well?
Why are particles or fields quantized in the first place?
\item What is an origin of particles (fields) and spacetime?
Which came first particles or space-time?
\item Why is our universe described by the standard model at the electroweak scale?
\end{enumerate}

As for Q1, the relationship between QFT and quantum mechanics (QM)
can be a key to solve the riddle by the following reasoning.
QM describes a system with a definite number of particles very well in a simple fashion.
In contrast, QFT is applied to a system that the number of particles can vary as well.
In other words, the range in application of QFT is wider than that of QM,
and QM can be actually reconstructed from QFT by fixing a number of particles
in a system.
Hence, if there were a theory that rebuilds QFT, an answer to Q1 can be derived.

As for Q2, this question can be expanded and deepened as 
``what is an origin of physical laws and our universe?''
and ``which came first physical laws or our universe?'', respectively.
If there were a framework to deal with 
particles and space-time (physical laws and our universe) as a unit,
the which-came-first-particles-or-spacetime problem 
(which-came-first-physical-laws-or-our-universe problem) could be solved.

As for Q3, this question stems from the fact that
QFT possesses no powerful principle to select realistic models 
theoretically and completely.
If there were a huge variety of universes with different particle contents
and physical parameters, called ``the level II multiverse''~\cite{T},
there is a possibility that the existence of our universe is 
understood by the anthropic principle~\cite{an1,an2},
and a profound riddle like the cosmological constant problem 
is neutralized~\cite{c-pr}.
Then, Q3 substitutes for the question whether a framework 
to describe the level II multiverse can be constructed or not.

In this paper, we explore a wider theoretical framework that has QFT built-in,
taking the fact that QM is reconstructed from QFT as a hint.
We formulate a quantum theory with an embedded structure
by introducing functional operators, 
and we find that it could describe the level II multiverse.
Topics related to a beginning of the universe such as an inflation, the third quantization
and the landscape are discussed in our formulation.

The outline of this paper is as follows.
In the next section, we review a framework of QFT
and explain how QM is derived from QFT.
In Sect. 3, we explore an underlying framework that embeds QFT, by the use of a toy model.
In Sect. 4, we extend our framework in order to describe the level II multiverse
and discuss physical implications on a birth of the universe.
In the last section, we give conclusions and discussions.
Explicit forms of Hamiltonian operators are listed
for several species of particle in appendix A.
More about a wave functional is explained in appendix B.

\section{Quantum field theory and quantum mechanics}

\subsection{Framework of quantum field theory}

First, we review a framework of QFT,
based on the Lagrangian density given by
\begin{eqnarray}
\widehat{\mathscr{L}}_{\varphi}
= \widehat{\varphi}^{\dagger}(\bm{x}, t) 
i\hbar \frac{\partial}{\partial t}\widehat{\varphi}(\bm{x}, t)
- \widehat{\varphi}^{\dagger}(\bm{x}, t) \widehat{H} \widehat{\varphi}(\bm{x}, t),
\label{L-varphi}
\end{eqnarray}
where $\widehat{\varphi}(\bm{x}, t)$ is a quantum field,
$\bm{x}=(x^1, x^2, x^3)$ and $t$ stand for coordinates of space and time, respectively,
$\hbar$ is the reduced Planck constant,
and $\widehat{H}$ is the Hamiltonian operator in QM containing $\bm{x}$ 
and its derivatives $\bm{\nabla}$, i.e.,
$\widehat{H}=\widehat{H}(\bm{x},-i\hbar\bm{\nabla})$.
Here, we choose the Lagrangian density with a first time-derivative term,
because the compatibility (relationship) with QM is easily comprehensible, as will be seen.

For example, for a particle with a mass $m$
in non-relativistic QM, $\widehat{H}$ is given by
\begin{eqnarray}
\widehat{H} = - \frac{\hbar^2}{2m} \bm{\nabla}^2 + V(\bm{x}),
\label{H-NRQM}
\end{eqnarray}
where $V(\bm{x})$ is a potential energy.
For a free Dirac fermion (a particle with spin $1/2$ and a mass $m$)
in relativistic QM,
$\widehat{H}$ is given by
\begin{eqnarray}
\widehat{H} = - i \hbar c \bm{\alpha} \cdot \bm{\nabla} + \beta mc^2 ,
\label{H-RQM}
\end{eqnarray}
where $c$ is the speed of light,
$\bm{\alpha}=(\alpha^1, \alpha^2, \alpha^3)$ 
and $\beta$ are $4 \times 4$ Hermitian matrices 
satisfying $\alpha^i\alpha^j+\alpha^j\alpha^i = 2 \delta^{ij}I$
($i,j=1,2,3$,$I$:$4 \times 4$ unit matrix),
$\alpha^i\beta + \beta\alpha^i=0$ and $\beta^2 = I$.

For simplicity, we assume that our spacetime is the four-dimensional
Minkowski spacetime and the system is described by a free field operator 
$\widehat{\varphi}(\bm{x}, t)$ in the Heisenberg picture.

The canonical conjugate of $\widehat{\varphi}(\bm{x}, t)$ is defined by
\begin{eqnarray}
\widehat{\pi}(\bm{x}, t) \equiv 
\frac{\partial \widehat{\mathscr{L}}_{\varphi}}
{\partial (\partial\widehat{\varphi}(\bm{x}, t)/\partial t)}
= i \hbar \widehat{\varphi}^{\dagger}(\bm{x}, t),
\label{pi}
\end{eqnarray}
and the Hamiltonian operator $\widehat{H}_{\varphi}$ in QFT is obtained as
\begin{align}
\widehat{H}_{\varphi} 
&\equiv \int \left(\widehat{\pi}(\bm{x}, t) 
\frac{\partial\widehat{\varphi}(\bm{x}, t)}{\partial t}
- \widehat{\mathscr{L}}_{\varphi}\right)d^3x
\nonumber \\
&= \int \widehat{\varphi}^{\dagger}(\bm{x}, t) 
\widehat{H} \widehat{\varphi}(\bm{x}, t) d^3x
= \frac{1}{i\hbar} \int \widehat{\pi}(\bm{x}, t) 
\widehat{H} \widehat{\varphi}(\bm{x}, t) d^3x.
\label{H-varphi}
\end{align}
We notice that the Hamiltonian operator in QFT is constructed by sandwiching
the Hamiltonian operator in QM between two field operators.
We refer to this construction as ``embedded structure'', ``nested construction''
or so on.
We list explicit forms of $\widehat{H}_{\varphi}$ 
for particles with spin 1/2 in appendix A.

The following quantization conditions are imposed on field operators with spin 1/2,
\begin{eqnarray}
\left\{\widehat{\varphi}(\bm{x}, t), \widehat{\pi}(\bm{y}, t)\right\} 
= i \hbar \delta^3(\bm{x} - \bm{y}),~~
\left\{\widehat{\varphi}(\bm{x}, t), \widehat{\varphi}(\bm{y}, t)\right\} = 0,~~
\left\{\widehat{\pi}(\bm{x}, t), \widehat{\pi}(\bm{y}, t)\right\} = 0,
\label{anti-c}
\end{eqnarray}
where $\{\widehat{A}, \widehat{B}\} \equiv \widehat{A}\widehat{B} + \widehat{B}\widehat{A}$.

Field operators obey the Heisenberg's equation of motion:
\begin{eqnarray}
i\hbar \frac{\partial}{\partial t}\widehat{\varphi}(\bm{x}, t) 
= \left[\widehat{\varphi}(\bm{x}, t), \widehat{H}_{\varphi}\right],~~
i\hbar \frac{\partial}{\partial t}\widehat{\pi}(\bm{x}, t) 
= \left[\widehat{\pi}(\bm{x}, t), \widehat{H}_{\varphi}\right],
\label{H-eqs}
\end{eqnarray}
where $[\widehat{A}, \widehat{B}] \equiv \widehat{A}\widehat{B} - \widehat{B}\widehat{A}$.
Using eqs.~\eqref{H-varphi}, \eqref{anti-c} and \eqref{H-eqs},
we derive the equations:
\begin{eqnarray}
i\hbar \frac{\partial}{\partial t}\widehat{\varphi}(\bm{x}, t) 
= \widehat{H} \widehat{\varphi}(\bm{x}, t),~~
i\hbar \frac{\partial}{\partial t}\widehat{\pi}(\bm{x}, t) 
= -\widehat{\pi}(\bm{x}, t) \widehat{H},
\label{H-eqs2}
\end{eqnarray}
and these equations agree with the Euler-Lagrange equation:
\begin{eqnarray}
\partial_{\mu}
\left(\frac{\partial \widehat{\mathscr{L}}_{\varphi}}
{\partial (\partial_{\mu}\widehat{\varphi}^{\dagger})}\right) 
- \frac{\partial \widehat{\mathscr{L}}_{\varphi}}
{\partial \widehat{\varphi}^{\dagger}}=0,~~
\partial_{\mu}
\left(\frac{\partial \widehat{\mathscr{L}}_{\varphi}}
{\partial (\partial_{\mu}\widehat{\varphi})}\right) 
- \frac{\partial \widehat{\mathscr{L}}_{\varphi}}
{\partial \widehat{\varphi}}=0,
\label{EL-eq}
\end{eqnarray}
that derived from the action integral 
$\displaystyle{\widehat{S}_{\varphi} 
= \frac{1}{c} \int \widehat{\mathscr{L}}_{\varphi} d^4x}$, 
based on a least action principle.

In the Schr\"{o}dinger picture, 
quantum fields $\widehat{\varphi}(\bm{x})$
and $\widehat{\pi}(\bm{x})$ are independent of time, 
and they are related to those in the Heisenberg picture as
\begin{eqnarray}
\widehat{\varphi}(\bm{x}, t) = e^{\frac{i}{\hbar}\widehat{H}_{\varphi}t}
\widehat{\varphi}(\bm{x}) e^{-\frac{i}{\hbar}\widehat{H}_{\varphi}t},~~
\widehat{\pi}(\bm{x}, t) = e^{\frac{i}{\hbar}\widehat{H}_{\varphi}t}
\widehat{\pi}(\bm{x}) e^{-\frac{i}{\hbar}\widehat{H}_{\varphi}t}.
\label{varphi-S}
\end{eqnarray}
Using eq.~\eqref{varphi-S} and the conservation law of $\widehat{H}_{\varphi}$,
i.e., ${d\widehat{H}_{\varphi}}/{dt}=0$,
$\widehat{H}_{\varphi}$ is rewritten 
in a time-independent form as
\begin{eqnarray}
\widehat{H}_{\varphi} 
= \int \widehat{\varphi}^{\dagger}(\bm{x}) 
\widehat{H} \widehat{\varphi}(\bm{x}) d^3x
= \frac{1}{i\hbar} \int \widehat{\pi}(\bm{x}) 
\widehat{H} \widehat{\varphi}(\bm{x}) d^3x.
\label{H-varphi-S}
\end{eqnarray}

Field operators $\widehat{\varphi}(\bm{x})$
and $\widehat{\pi}(\bm{x})$ obey the anti-commutation relations:
\begin{eqnarray}
\left\{\widehat{\varphi}(\bm{x}), \widehat{\pi}(\bm{y})\right\} 
= i \hbar \delta^3(\bm{x} - \bm{y}),~~
\left\{\widehat{\varphi}(\bm{x}), \widehat{\varphi}(\bm{y})\right\} = 0,~~
\left\{\widehat{\pi}(\bm{x}), \widehat{\pi}(\bm{y})\right\} = 0.
\label{anti-c-S}
\end{eqnarray}
From the first condition in eq.~\eqref{anti-c-S},
$\widehat{\pi}(\bm{y})$ is given by
$\widehat{\pi}(\bm{y}) = i\hbar {\delta}/{\delta {\varphi}(\bm{y})}$,
i.e., $\widehat{\varphi}^{\dagger}(\bm{y}) 
= {\delta}/{\delta {\varphi}(\bm{y})}$,
using the functional derivative,
in the representative diagonalizing $\widehat{\varphi}(\bm{x})$
such as $\widehat{\varphi}(\bm{x})|\varphi\rangle = {\varphi}(\bm{x})|\varphi\rangle$
where ${\varphi}(\bm{x})$ is a Grassmann-valued field configuration.

Any state $|\varPsi(t)\rangle$ is evolved by the Schr\"{o}dinger equation:
\begin{eqnarray}
i \hbar \frac{d}{dt} |\varPsi(t)\rangle = \widehat{H}_{\varphi} |\varPsi(t)\rangle,
\label{S-eq}
\end{eqnarray}
and its formal solution is given by
\begin{eqnarray}
|\varPsi(t)\rangle = e^{-\frac{i}{\hbar}\widehat{H}_{\varphi}t} |\varPsi(0)\rangle.
\label{S-eq-sol}
\end{eqnarray}
Multiplying $\langle \varphi|$ by both sides of eq.~(\ref{S-eq}),
we obtain the equation:
\begin{eqnarray}
i \hbar \frac{\partial}{\partial t} \varPsi(\varphi, t)
= \widehat{H}_{\varphi} \varPsi(\varphi, t),
\label{S-eq-QFT}
\end{eqnarray}
where $\varPsi(\varphi, t)=\langle \varphi|\varPsi(t)\rangle$ 
is a wave functional in QFT~\cite{Wf},
and $\widehat{H}_{\varphi}$ is written by
\begin{eqnarray}
\widehat{H}_{\varphi} 
= \int \frac{\delta}{\delta {\varphi}(\bm{x})} \widehat{H} {\varphi}(\bm{x}) d^3x,
\label{H-varphi-S-der}
\end{eqnarray}
using $\widehat{\pi}(\bm{x}) = i\hbar {\delta}/{\delta {\varphi}(\bm{x})}$.
The expectation value of $\widehat{H}_{\varphi}$ is given by
\begin{eqnarray}
\langle\varPsi(t)|\widehat{H}_{\varphi}|\varPsi(t)\rangle
= \int \mathscr{D}\varphi~
\varPsi^{\dagger}(\varphi, t)\widehat{H}_{\varphi}\varPsi(\varphi, t),
\label{<H>-QFT}
\end{eqnarray}
as seen in eq.~\eqref{B-<H>}.
We explain more about a wave functional in appendix B.

Here, for the sake of completeness, we comment on a boson with spin 0.
For a free complex scalar particle $\phi$ with a mass $m$,
the Lagrangian density in QFT is given by
\begin{eqnarray}
\widehat{\mathscr{L}}_{\phi}
= \hbar c \left\{\partial_{\mu}\widehat{\phi}^{\dagger}(\bm{x}, t) 
\partial^{\mu}\widehat{\phi}(\bm{x}, t)
- \left(\frac{mc}{\hbar}\right)^2\widehat{\phi}^{\dagger}(\bm{x}, t)\widehat{\phi}(\bm{x}, t)
\right\},
\label{L-phi}
\end{eqnarray}
where $\widehat{\phi}(\bm{x}, t)$ is the quantum field of $\phi$.
The canonical conjugate of $\widehat{\phi}(\bm{x}, t)$ 
and $\widehat{\phi}^{\dagger}(\bm{x}, t)$ are defined by
\begin{eqnarray}
\widehat{\pi}_{\phi}(\bm{x}, t) \equiv 
\frac{\partial \widehat{\mathscr{L}}_{\phi}}
{\partial (\partial\widehat{\phi}(\bm{x}, t)/\partial t)}
= \frac{\hbar}{c} \frac{\partial}{\partial t}\widehat{\phi}^{\dagger}(\bm{x}, t),~~
\widehat{\pi}^{\dagger}_{\phi}(\bm{x}, t) \equiv 
\frac{\partial \widehat{\mathscr{L}}_{\phi}}
{\partial (\partial\widehat{\phi}^{\dagger}(\bm{x}, t)/\partial t)}
= \frac{\hbar}{c} \frac{\partial}{\partial t}\widehat{\phi}(\bm{x}, t),
\label{pi-phi}
\end{eqnarray}
respectively, and the Hamiltonian operator $\widehat{H}_{\phi}$ is obtained as
\begin{align}
\widehat{H}_{\phi} 
= \int \hbar c \left\{\frac{1}{\hbar^2} 
\widehat{\pi}^{\dagger}_{\phi}(\bm{x}, t) \widehat{\pi}_{\phi}(\bm{x}, t)
+ \bm{\nabla}\widehat{\phi}^{\dagger}(\bm{x}, t)\cdot\bm{\nabla}\widehat{\phi}(\bm{x}, t)
+ \left(\frac{mc}{\hbar}\right)^2\widehat{\phi}^{\dagger}(\bm{x}, t)\widehat{\phi}(\bm{x}, t)
\right\}d^3x.
\label{H-phi}
\end{align}
The following commutation relations are imposed on the field operators with spin 0:
\begin{eqnarray}
\hspace{-1cm}
&~& \left[\widehat{\phi}(\bm{x}, t), \widehat{\pi}_{\phi}(\bm{y}, t)\right] 
= i \hbar \delta^3(\bm{x} - \bm{y}),~~
\left[\widehat{\phi}(\bm{x}, t), \widehat{\phi}(\bm{y}, t)\right] = 0,~~
\left[\widehat{\pi}_{\phi}(\bm{x}, t), \widehat{\pi}_{\phi}(\bm{y}, t)\right] = 0,
\nonumber \\
\hspace{-1cm}
&~& \left[\widehat{\phi}^{\dagger}(\bm{x}, t), \widehat{\pi}^{\dagger}_{\phi}(\bm{y}, t)\right] 
= i \hbar \delta^3(\bm{x} - \bm{y}),~~
\left[\widehat{\phi}^{\dagger}(\bm{x}, t), \widehat{\phi}^{\dagger}(\bm{y}, t)\right] = 0,~~
\left[\widehat{\pi}^{\dagger}_{\phi}(\bm{x}, t), 
\widehat{\pi}^{\dagger}_{\phi}(\bm{y}, t)\right] = 0,
\nonumber \\
\hspace{-1cm}
&~& \left[\widehat{\phi}(\bm{x}, t), \widehat{\pi}^{\dagger}_{\phi}(\bm{y}, t)\right] = 0,~~
\left[\widehat{\phi}^{\dagger}(\bm{x}, t), \widehat{\pi}_{\phi}(\bm{y}, t)\right] = 0,~~
\left[\widehat{\phi}(\bm{x}, t), \widehat{\phi}^{\dagger}(\bm{y}, t)\right] = 0,
\nonumber \\
\hspace{-1cm}
&~& \left[\widehat{\pi}_{\phi}(\bm{x}, t), \widehat{\pi}^{\dagger}_{\phi}(\bm{y}, t)\right] = 0.
\label{c-phi}
\end{eqnarray}
In the Schr\"{o}dinger picture, quantum fields are independent of time, and 
$\widehat{\pi}_{\phi}(\bm{y})$ and $\widehat{\pi}^{\dagger}_{\phi}(\bm{y})$ 
are represented by the functional derivatives such as
$\widehat{\pi}_{\phi}(\bm{y}) = - i\hbar {\delta}/{\delta {\phi}(\bm{y})}$
and $\widehat{\pi}^{\dagger}_{\phi}(\bm{y}) 
= - i\hbar {\delta}/{\delta {\phi}^{\dagger}(\bm{y})}$, respectively.
A wave functional $\varPsi(\phi, t)=\langle \phi|\varPsi(t)\rangle$
is evolved by the Schr\"{o}dinger equation:
\begin{eqnarray}
i \hbar \frac{\partial}{\partial t} \varPsi(\phi, t)
= \widehat{H}_{\phi} \varPsi(\phi, t),
\label{S-eq-phi-QFT}
\end{eqnarray}
where the Hamiltonian operator is rewritten as
\begin{align}
\widehat{H}_{\phi} 
= \int \hbar c \left\{-\frac{\delta^2}{\delta {\phi}^{\dagger}(\bm{x})\delta {\phi}(\bm{x})}
+ \bm{\nabla}{\phi}^{\dagger}(\bm{x})\cdot\bm{\nabla}{\phi}(\bm{x})
+ \left(\frac{mc}{\hbar}\right)^2{\phi}^{\dagger}(\bm{x}){\phi}(\bm{x})\right\}d^3x.
\label{H-phi-S}
\end{align}

In a similar way as a particle with spin 1/2 and a particle with spin 0,
a particle with spin 1 (gauge boson) and a particle with a higher spin 
are also described by the use of QFT.
Actually, particle physics at the electroweak scale is excellently 
controlled by the Lagrangian density $\widehat{\mathscr{L}}_{\rm SM}$ of the standard model,
and the basic formula in the Schr\"{o}dinger picture is given by
\begin{eqnarray}
i \hbar \frac{\partial}{\partial t} \varPsi(\varphi_{\rm SM}, t)
= \widehat{H}_{\rm SM} \varPsi(\varphi_{\rm SM}, t),
\label{S-eq-SM-QFT}
\end{eqnarray}
where $\widehat{H}_{\rm SM}$ is the Hamiltonian operator 
derived from $\widehat{\mathscr{L}}_{\rm SM}$,
and $\varphi_{\rm SM}$ denotes a set of field variables of the standard model particles.

Let us come back to the scalar particle $\phi$.
When we deal with a relativistic scalar particle in QM,
we encounter the problem that the probability interpretation breaks down
in the absence of a positive definite expression of probability density.
Because QM shows excellence in a description of a system with a definite number of particles,
we do well to consider a non-relativistic case.
Using the redefinition of quantum field:
\begin{eqnarray}
\widehat{\phi}(\bm{x}, t) 
= \sqrt{\frac{\hbar}{2mc}} e^{-\frac{i}{\hbar}mc^2 t}~\widehat{\psi}(\bm{x}, t),
\label{field-redef}
\end{eqnarray}
$\widehat{\mathscr{L}}_{\phi}$ is rewritten as
\begin{align}
\widehat{\mathscr{L}}_{\phi}
&= \frac{i\hbar}{2}\left(\widehat{\psi}^{\dagger}(\bm{x}, t) 
\frac{\partial}{\partial t} \widehat{\psi}(\bm{x}, t)
- \frac{\partial}{\partial t} \widehat{\psi}^{\dagger}(\bm{x}, t)~ 
\widehat{\psi}(\bm{x}, t)\right)
- \frac{\hbar^2}{2m} \bm{\nabla}\widehat{\psi}^{\dagger}(\bm{x}, t)\cdot
\bm{\nabla}\widehat{\psi}(\bm{x}, t) + \cdots
\nonumber \\
&= \widehat{\psi}^{\dagger}(\bm{x}, t) 
i\hbar \frac{\partial}{\partial t} \widehat{\psi}(\bm{x}, t)
- \widehat{\psi}^{\dagger}(\bm{x}, t) \left(-\frac{\hbar^2}{2m} \bm{\nabla}^2\right)
\widehat{\psi}(\bm{x}, t) + ({\rm total~~derivatives}) + \cdots,
\label{L-psi}
\end{align}
where the ellipsis contains a second time-derivative term 
and this term is neglected in the non-relativistic limit, i.e.,  
$\displaystyle{|mc^2 \widehat{\psi}| \gg |i\hbar\partial\widehat{\psi}}/{\partial t}|$.
From eq.~\eqref{L-psi}, we find that 
the Hamiltonian operator has an embedded structure such as
$\displaystyle{\widehat{H}_{\psi} 
= \int \widehat{\psi}^{\dagger}(\bm{x}, t)\widehat{H} \widehat{\psi}(\bm{x}, t)d^3x}$ 
with $\displaystyle{\widehat{H} = -\frac{\hbar^2}{2m} \bm{\nabla}^2}$.

\subsection{Derivation of quantum mechanics}

Let us examine the relationship between QFT and QM,
paying attention to physical states, operators of four-momenta and those expectation values.

A physical state in QFT is represented in $\bm{x}$-representation
($\bm{x}$-space) as a superposition of states 
with various numbers of particles as follows,\footnote{
Strictly speaking, we need to include anti-particles, but we omit them for simplicity.}
\begin{eqnarray}
|\varPsi(t)\rangle 
= \psi^{(0)}(t) |0 \rangle + 
\sum_{N = 1}^{\infty} \int d^3x_1 \cdots d^3x_N~
\psi(\bm{x}_1, \cdots, \bm{x}_N, t) |\bm{x}_1, \cdots, \bm{x}_N\rangle,
\label{state-QFT}
\end{eqnarray}
where $|0\rangle$ is a vacuum state in QFT, 
$\psi(\bm{x}_1, \cdots, \bm{x}_N, t)$ is a wave function
of $N$-particle state in QM defined by
\begin{eqnarray}
\psi(\bm{x}_1, \cdots, \bm{x}_N, t) 
\equiv \langle\bm{x}_1, \cdots, \bm{x}_N|\varPsi(t)\rangle,~~
\label{psi}
\end{eqnarray}
and $|\bm{x}_1, \cdots, \bm{x}_N\rangle$
is the ket vector which satisfies
$\widehat{\bm{x}}_l |\bm{x}_1, \cdots, \bm{x}_N\rangle
= \bm{x}_l|\bm{x}_1, \cdots, \bm{x}_N\rangle$ ($l=1, \cdots, N$)
and is defined by
\begin{eqnarray}
|\bm{x}_1, \cdots, \bm{x}_N\rangle
\equiv \frac{1}{\sqrt{N!}} \widehat{\varphi}^{\dagger}(\bm{x}_1) 
\cdots \widehat{\varphi}^{\dagger}(\bm{x}_N) 
|0\rangle,
\label{x-state}
\end{eqnarray}
using field operators $\widehat{\varphi}^{\dagger}(\bm{x})$.
Note that $\widehat{\varphi}^{\dagger}(\bm{x})$ plays a role of 
a creation operator of a particle $\varphi$ and
an annihilation operator of its anti-particle $\overline{\varphi}$,
and $\widehat{\varphi}(\bm{x})$ plays a role of 
a creation operator of $\overline{\varphi}$ and
an annihilation operator of $\varphi$.

First, we consider a zero-particle state (a state of vacuum) such as
\begin{eqnarray}
|\varPsi(t)\rangle^{(0)} = \psi^{(0)}(t)|0\rangle,
\label{state-0-QM}
\end{eqnarray}
where $\psi^{(0)}(t)$ is a wave function of zero-particle state in QM
and is rewritten by $\psi^{(0)}(t) = \langle 0|\varPsi(t)\rangle$.
The $\psi^{(0)}(t)$ satisfies the equation:
\begin{eqnarray}
i \hbar \frac{d}{dt} \psi^{(0)}(t)
= \mathscr{E}_0\psi^{(0)}(t),
\label{S-eq-0-QM}
\end{eqnarray}
and its solution is obtained as 
$\psi^{(0)}(t) = e^{-\frac{i}{\hbar}\mathscr{E}_0 t}\psi^{(0)}(0)$
with a vacuum energy $\mathscr{E}_0=\langle 0|\widehat{H}_{\varphi}|0\rangle$.

Next we consider one-particle state limited as
\begin{eqnarray}
|\varPsi(t)\rangle^{(1)} = |\psi(t)\rangle = \int d^3x~\psi(\bm{x}, t) |\bm{x}\rangle,
\label{state-1-QM}
\end{eqnarray}
where $\psi(\bm{x}, t)$ is a wave function of one-particle state in QM
and is rewritten by
\begin{align}
\psi(\bm{x}, t)&=\langle\bm{x}|\varPsi(t)\rangle 
=\langle 0|\widehat{\varphi}(\bm{x})|\varPsi(t)\rangle
=\langle 0|e^{\frac{i}{\hbar}\widehat{H}_{\varphi}t}\widehat{\varphi}(\bm{x})
e^{-\frac{i}{\hbar}\widehat{H}_{\varphi}t}|\varPsi(0)\rangle 
\nonumber \\ 
&=\langle 0|\widehat{\varphi}(\bm{x},t)|\varPsi(0)\rangle,
\label{psi-1}
\end{align}
using eq.~\eqref{S-eq-sol}, 
$\langle 0|e^{\frac{i}{\hbar}\widehat{H}_{\varphi}t}=\langle 0|$, i.e.,
$\langle 0|\widehat{H}_{\varphi} = 0$ and eq.~\eqref{varphi-S}.
Here, according to the ordinary procedure,
we redefine $\widehat{H}_{\varphi}$ by using the normal ordering.
From eq.~\eqref{psi-1} and the first equation in eq.~\eqref{H-eqs2},
we find that $\psi(\bm{x}, t)$ satisfies the Schr\"{o}dinger equation in QM:
\begin{eqnarray}
i \hbar \frac{\partial}{\partial t} \psi(\bm{x}, t)
= \widehat{H} \psi(\bm{x}, t)
\label{S-eq-QM}
\end{eqnarray}
with the Hamiltonian operator
$\widehat{H}=\widehat{H}(\bm{x},-i\hbar\bm{\nabla})$ in QM.
The Schr\"{o}dinger equation is rewritten by
\begin{eqnarray}
i \hbar \frac{d}{dt} |\psi(t)\rangle
= \widehat{H}|\psi(t)\rangle,
\label{S-eq-QM-state}
\end{eqnarray}
and its formal solution is given by
\begin{eqnarray}
|\psi(t)\rangle = e^{-\frac{i}{\hbar}\widehat{H}t} |\psi(0)\rangle.
\label{S-eq-QM-sol}
\end{eqnarray}

In a similar way, using eq.~\eqref{S-eq}, 
we find that $\psi(\bm{x}_1, \cdots, \bm{x}_N, t)$ obeys the equation:
\begin{eqnarray}
i \hbar \frac{\partial}{\partial t} \psi(\bm{x}_1, \cdots, \bm{x}_N, t)
= \widehat{H}^{(N)}\psi(\bm{x}_1, \cdots, \bm{x}_N, t),
\label{S-eq-N-QM}
\end{eqnarray}
where $\widehat{H}^{(N)}$ is a total Hamiltonian operator of non-interacting $N$ particles:
\begin{eqnarray}
\widehat{H}^{(N)} = \sum_{l=1}^{N} \widehat{H}(\bm{x}_l, -i\hbar\bm{\nabla}_l).
\label{H-N}
\end{eqnarray}

In a quantum theory, four-momenta become generators of
the spacetime translation such as $x^{\mu} \to x'^{\mu}=x^{\mu} - \varepsilon^{\mu}$ 
where $\mu = 0, 1, 2, 3$, $x^0 = ct$, $\bm{x}=(x^1, x^2, x^3)$ 
and $\varepsilon^{\mu}$ is an infinitesimal constant four-vector.
Using the Noether procedure in analytical mechanics of fields,
we derive the operators of four-momenta $\widehat{P}_{\mu}$ in QFT:
\begin{eqnarray}
\widehat{P}_{\mu} 
= \frac{1}{c} \int 
\left(\frac{\partial \widehat{\mathscr{L}}_{\varphi}}{\partial (\partial_{0}\widehat{\varphi})}
\partial_{\mu}\widehat{\varphi} 
- {\delta^{0}}_{\mu} \widehat{\mathscr{L}}_{\varphi}\right)d^3x
= \frac{1}{c} \int {\widehat{T}^0}_{~\mu}d^3x,
\label{Pmu}
\end{eqnarray}
where $\partial_0 \widehat{\varphi} 
= {\partial \widehat{\varphi}}/{\partial x^0}$,
${\widehat{T}^0}_{~\mu}$ is a component of the energy-momentum tensor,
and we use the fact that the field operator 
$\widehat{\varphi}(x) (=\widehat{\varphi}(x^0, \bm{x}))$ transforms as 
$\widehat{\varphi}'(x') = \widehat{\varphi}'(x-\varepsilon)
=\widehat{\varphi}(x)$ under the translation $x'^{\mu}=x^{\mu} - \varepsilon^{\mu}$
and a change of $\widehat{\varphi}(x)$ is given by
\begin{eqnarray}
\delta_{\varepsilon} \widehat{\varphi}(x) 
= \widehat{\varphi}'(x) - \widehat{\varphi}(x)
= \widehat{\varphi}(x+\varepsilon) - \widehat{\varphi}(x)
= \varepsilon^{\mu} \partial_{\mu} \widehat{\varphi}(x).
\label{delta-varphi}
\end{eqnarray}

Because the Minkowski spacetime has a homogeneity,
a physical system is invariant under the translation
and then four-momenta conserve, 
i.e., $i\hbar{d\widehat{P}_{\mu}}/{dt} 
=[\widehat{P}_{\mu}, \widehat{H}] = 0$.
Using eqs.~\eqref{L-varphi}, \eqref{Pmu} and the conservation law of $\widehat{P}_{\mu}$, 
we obtain the formula of $\widehat{P}_{\mu}$:
\begin{eqnarray}
\hspace{-1cm}
&~& \widehat{P}_0 = \frac{1}{c} \int \widehat{\varphi}^{\dagger}(x) 
\widehat{H}(\bm{x},-i\hbar\bm{\nabla}) \widehat{\varphi}(x) d^3x 
= \frac{1}{c} \int \widehat{\varphi}^{\dagger}(\bm{x}) 
\widehat{H}(\bm{x},-i\hbar\bm{\nabla}) \widehat{\varphi}(\bm{x}) d^3x
= \frac{\widehat{H}_{\varphi}}{c},
\label{P0}\\
\hspace{-1cm}
&~& \widehat{\bm{P}} = \int \widehat{\varphi}^{\dagger}(x) 
\left(-i \hbar \bm{\nabla}\right) \widehat{\varphi}(x) d^3x
= \int \widehat{\varphi}^{\dagger}(\bm{x}) 
\left(-i \hbar \bm{\nabla}\right) \widehat{\varphi}(\bm{x}) d^3x.
\label{bmP}
\end{eqnarray}
In a similar way, we obtain the formula of orbital angular momenta $\widehat{\bm{J}}$:
\begin{eqnarray}
\widehat{\bm{J}} = \int \widehat{\varphi}^{\dagger}(\bm{x}) 
\left\{\bm{x} \times \left(-i \hbar \bm{\nabla}\right)\right\} 
\widehat{\varphi}(\bm{x}) d^3x,
\label{J}
\end{eqnarray}
from the rotation.
In this way, it is ascertained that a nested structure exists
for $\widehat{{\bm P}}$ and $\widehat{\bm{J}}$
other than the Hamiltonian operator $c\widehat{P}_{0}$, 
as seen from the formulas \eqref{bmP} and \eqref{J}.

Or, under the assumption that a nesting structure is present
such as eqs.~(\ref{bmP}) and (\ref{J}), we reaffirm that the momenta are
represented by the differential operator $\widehat{\bm{p}} = -i\hbar \bm{\nabla}$
in $\bm{x}$-representation
and the following commutation relations in QM holds:
\begin{eqnarray}
[\widehat{x}^i, \widehat{p}^j] = i\hbar \delta^{ij},~~ 
[\widehat{x}^i, \widehat{x}^j] = 0,~~ [\widehat{p}^i, \widehat{p}^j] = 0.
\label{[x,p]}
\end{eqnarray}

Let us evaluate expectation values of momenta $\bm{P}$ in the momentum space.
There, a physical state in QFT is represented by
\begin{eqnarray}
|\varPsi(t)\rangle 
= \psi^{(0)}(t) |0 \rangle + 
\sum_{N = 1}^{\infty} \int d^3k_1 \cdots d^3k_N~
\widetilde{\psi}(\bm{k}_1, \cdots, \bm{k}_N, t) |\bm{k}_1, \cdots, \bm{k}_N\rangle,
\label{state-p-QFT}
\end{eqnarray}
where $\widetilde{\psi}(\bm{k}_1, \cdots, \bm{k}_N, t)$ is a wave function
of $N$-particle state in QM, $\bm{k}_l$ ($l=1, \cdots, N$) are wave-number vectors
and $|\bm{k}_1, \cdots, \bm{k}_N\rangle$ is the ket vector defined by
\begin{eqnarray}
|\bm{k}_1, \cdots, \bm{k}_N\rangle
\equiv \frac{1}{\sqrt{N!}} \widehat{b}^{\dagger}(\bm{k}_1) 
\cdots \widehat{b}^{\dagger}(\bm{k}_N) 
|0\rangle,
\label{p-state}
\end{eqnarray}
using creation operators $\widehat{b}^{\dagger}(\bm{k}_l)$ of particle $\varphi$
with the momenta $\bm{p}_l=\hbar \bm{k}_l$.

The momentum operator in QFT is written as
\begin{eqnarray}
\widehat{\bm{P}}
= \int \hbar\bm{k} \left(\widehat{b}^{\dagger}(\bm{k}) \widehat{b}(\bm{k}) 
+ \widehat{d}^{\dagger}(\bm{k}) \widehat{d}(\bm{k})\right)d^3k,
\label{P-p-S}
\end{eqnarray}
where $\widehat{b}(\bm{k})$ is the annihilation operator of $\varphi$ with 
the momenta $\bm{p}=\hbar \bm{k}$,
and $\widehat{d}^{\dagger}(\bm{k})$ and $\widehat{d}(\bm{k})$ are 
the creation and annihilation operators of 
the anti-particle $\overline{\varphi}$ with $\bm{p}=\hbar \bm{k}$, respectively.

For a one-particle state given by
\begin{eqnarray}
|\varPsi(t)\rangle^{(1)} = |\psi(t)\rangle = \int d^3k~\widetilde{\psi}(\bm{k}, t) 
\widehat{b}^{\dagger}(\bm{k})|0\rangle,
\label{state-1-p-QM}
\end{eqnarray}
the expectation value of $\bm{P}$ is calculated as
\begin{align}
^{(1)}\langle \varPsi(t)|\widehat{\bm{P}}|\varPsi(t)\rangle^{(1)}
= \langle \psi(t)|\widehat{\bm{P}}|\psi(t)\rangle
= \int d^3k~\tilde{\psi}^{\dagger}(\bm{k},t)~\hbar\bm{k}~\widehat{\psi}(\bm{k}, t)
= \langle \psi(t)|\widehat{\bm{p}}|\psi(t)\rangle,
\label{<p>-p-QM}
\end{align}
where we use 
$\{\widehat{b}(\bm{k}), \widehat{b}^{\dagger}(\bm{k}')\}=\delta^3(\bm{k}-\bm{k}')$,
$\widehat{b}(\bm{k})| 0 \rangle = 0$, $\widehat{d}(\bm{k})| 0 \rangle = 0$,
$\langle 0 | 0 \rangle = 1$ and so on.
Together with the normalization condition:
\begin{eqnarray}
^{(1)}\langle \varPsi(t)|\varPsi(t)\rangle^{(1)}
= \langle \psi(t)|\psi(t)\rangle
= \int d^3k~\widetilde{\psi}^{\dagger}(\bm{k},t) \widetilde{\psi}(\bm{k}, t) = 1,
\label{<I>-QM}
\end{eqnarray}
we arrive at the probability interpretation that
$\widetilde{\psi}^{\dagger}(\bm{k},t) \widetilde{\psi}(\bm{k}, t)$
is the probability density and 
$\langle \psi(t)|\widehat{\bm{p}}|\psi(t)\rangle$
is the expectation value of momenta for the one-particle system in QM.

In a similar way, for a non-interacting $N$-particle system, 
the expectation value of total momenta is given by
\begin{eqnarray}
^{(N)}\langle \varPsi(t)|\widehat{\bm{P}}|\varPsi(t)\rangle^{(N)}
= \int d^3k_1 \cdots d^3k_N~
\widetilde{\psi}^{\dagger}(\bm{k}_1, \cdots, \bm{k}_N, t) 
\sum_{l=1}^{N} \hbar\bm{k}_l
\widetilde{\psi}(\bm{k}_1, \cdots, \bm{k}_N, t),
\label{<P>-N-p-QM}
\end{eqnarray}
where $|\varPsi(t)\rangle^{(N)}$ is a state vector for the $N$-particle system
given by
\begin{eqnarray}
|\varPsi(t)\rangle^{(N)} = \int d^3k_1 \cdots d^3k_N~
\widetilde{\psi}(\bm{k}_1, \cdots, \bm{k}_N, t) |\bm{k}_1, \cdots, \bm{k}_N\rangle.
\label{state-N-p-QM}
\end{eqnarray}

In the Sch\"odinger picture, observables in a system described by a Lagrangian density
with a first time-derivative term are, in general, given as a form
\begin{eqnarray}
\widehat{\varOmega}_{\varphi}^{a} 
= \int \widehat{\varphi}^{\dagger}(\bm{x}) 
\widehat{\varOmega}^a \widehat{\varphi}(\bm{x}) d^3x,
\label{Omega}
\end{eqnarray}
where $\widehat{\varOmega}_{\varphi}^{a}$ are operators including quantum fields in QFT,
and $\widehat{\varOmega}^a=\widehat{\varOmega}^a(\bm{x}, -i\hbar\bm{\nabla})$ 
are operators operating a wave function in the position space of QM.
It is shown that both $\widehat{\varOmega}_{\varphi}^{a}$ 
and $\widehat{\varOmega}^a$
satisfy the same type of algebraic relations:
\begin{eqnarray}
\left[\widehat{\varOmega}_{\varphi}^{a}, \widehat{\varOmega}_{\varphi}^{b}\right]
= i \sum_{c} f^{abc} \widehat{\varOmega}_{\varphi}^{c},~~
\left[\widehat{\varOmega}^{a}, \widehat{\varOmega}^{b}\right]
= i \sum_{c} f^{abc} \widehat{\varOmega}^{c}.
\label{Omega-c}
\end{eqnarray}

Changing the Sch\"odinger picture into the Heisenberg picture,
$\widehat{\bm{x}}$ and $\widehat{\bm{p}}$ 
possess time-dependence such as
\begin{eqnarray} 
\widehat{\bm{x}}(t)=e^{\frac{i}{\hbar}\widehat{H}t}\widehat{\bm{x}}
e^{-\frac{i}{\hbar}\widehat{H}t},~~
\widehat{\bm{p}}(t)=e^{\frac{i}{\hbar}\widehat{H}t}\widehat{\bm{p}}
e^{-\frac{i}{\hbar}\widehat{H}t},
\label{x(t)p(t)}
\end{eqnarray}
and $\widehat{\bm{x}}(t)$ and $\widehat{\bm{p}}(t)$ obey the Heisenberg's equation of motion:
\begin{eqnarray}
i\hbar\frac{d\widehat{\bm{x}}(t)}{dt} = [\widehat{\bm{x}}(t), \widehat{H}],~~ 
i\hbar\frac{d\widehat{\bm{p}}(t)}{dt} = [\widehat{\bm{p}}(t), \widehat{H}].
\label{H-eq-QM}
\end{eqnarray}

Here, for the sake of completeness,
we point out that QM is also reconstructed from QFT
in the case that particles interact among them.
In concrete, a Hamiltonian operator consists of two parts such as
$\widehat{H}_{\varphi} = \widehat{H}_{\varphi}^{(0)} + \widehat{H}_{\varphi}^{\rm int}$
where $\widehat{H}_{\varphi}^{(0)}$ is a part relating to a kinetic energy
and $\widehat{H}_{\varphi}^{\rm int}$ represents interaction among particles.
In the interaction picture, field operators $\widehat{\varphi}_{\rm I}(\bm{x}, t)$
and $\widehat{\pi}_{\rm I}(\bm{x}, t)$ obey the Heisenberg's equation of motion:
\begin{eqnarray}
i\hbar \frac{\partial}{\partial t}\widehat{\varphi}_{\rm I}(\bm{x}, t) 
= \left[\widehat{\varphi}_{\rm I}(\bm{x}, t), \widehat{H}_{\varphi}^{(0)}\right],~~
i\hbar \frac{\partial}{\partial t}\widehat{\pi}_{\rm I}(\bm{x}, t) 
= \left[\widehat{\pi}_{\rm I}(\bm{x}, t), \widehat{H}_{\varphi}^{(0)}\right],
\label{H-eqs-int}
\end{eqnarray}
they behave as free fields,
and the Fock space is constructed using them.
The physical state $|\varPsi_{\rm I}(t)\rangle$ 
is evolved by the Schr\"{o}dinger equation:
\begin{eqnarray}
i \hbar \frac{d}{dt} |\varPsi_{\rm I}(t)\rangle 
= \widehat{H}_{\varphi}^{\rm int}(\widehat{\varphi}_{\rm I},
\widehat{\pi}_{\rm I}) |\varPsi_{\rm I}(t)\rangle,
\label{S-eq-int}
\end{eqnarray}
and, for a one-particle state, 
we can effectively derive the Schr\"{o}dinger equation
in the interaction picture of QM:
\begin{eqnarray}
i \hbar \frac{d}{dt} |\psi_{\rm I}(t)\rangle 
= \widehat{V}_{\rm I}(t)|\psi_{\rm I}(t)\rangle
\label{S-eq-int-QM}
\end{eqnarray}
with a potential $\widehat{V}_{\rm I}(t)$, using the Born approximation
in the non-relativistic limit.

We list central features concerning 
the relationship between QFT and QM, in a system described by a Lagrangian density
with a first time-derivative term.
\begin{itemize}
\item QM is reconstructed from QFT under the condition
that a number of particles is unchanged.
\item From a transformation property of quantum fields 
under the translation and the rotation, it is understood that the momenta $\bm{p}$ are
represented by the differential operator $\widehat{\bm{p}} = -i\hbar \bm{\nabla}$
in $\bm{x}$-representation of QM.
\item Observables $\widehat{\varOmega}_{\varphi}^{a}$ in QFT are, in general, constructed 
in the form of an embedded structure, e.g.,
$\displaystyle{\widehat{\varOmega}_{\varphi}^{a} 
= \int \widehat{\varphi}^{\dagger}(\bm{x}) 
\widehat{\varOmega}^a \widehat{\varphi}(\bm{x}) d^3x}$,
where $\widehat{\varOmega}^a=\widehat{\varOmega}^a(\bm{x}, -i\hbar\bm{\nabla})$ 
are operators operating a wave function in $\bm{x}$-space of QM.
\end{itemize}

\section{Quantum field's functional theory}

Let us come back to Q1, i.e., ``Why does QFT work as an effective theory 
of elementary particles extremely well?~
Why are particles or fields quantized in the first place?''.
In concrete, why is a fermion (boson) described by the field operator such as
$\widehat{\pi}(\bm{x}) = i\hbar {\delta}/{\delta {\varphi}(\bm{x})}$,
i.e., $\widehat{\varphi}^{\dagger}(\bm{x}) 
= {\delta}/{\delta {\varphi}(\bm{x})}$
($\widehat{\pi}(\bm{x}) = -i\hbar {\delta}/{\delta {\varphi}(\bm{x})}$),
using the functional derivative?

If there were a fundamental framework 
that reaches QFT in a similar way as the derivation of QM from QFT,
we can have an answer to the above questions.
To explore such a framework,
we will draw on the embedded structure in QFT, e.g.,
\begin{eqnarray}
\widehat{H}_{\varphi} = \int \widehat{\varphi}^{\dagger}(x) 
\widehat{H}\widehat{\varphi}(x) d^3x
= \frac{1}{i\hbar} \int \widehat{\pi}(x) 
\widehat{H} \widehat{\varphi}(x) d^3x.
\label{H-nest}
\end{eqnarray}
For simplicity, we consider a toy model containing a particle $\varphi$ alone in this section.

\subsection{Framework of quantum field's functional theory}

Taking a hint from the nested construction \eqref{H-nest}, 
we introduce basic operators 
denoted as $\widehat{\varPhi}(\{\varphi\}, t)$ 
and its canonical conjugate $\widehat{\varPi}(\{\varphi\}, t)$,
whose role will be found out in the next subsection.
We refer to $\widehat{\varPhi}(\{\varphi\}, t)$ and $\widehat{\varPi}(\{\varphi\}, t)$
as ``functional operators'' or ``field's functional''.
And we call a theory of functional operator
``quantum field's functional theory'',
``quantum theory of field's functionals'' or ``QFFT'' in short.

Let us start with the Lagrangian owning a nested structure:
\begin{eqnarray}
\widehat{{L}}_{\varPhi}
= \widehat{\varPhi}^{\dagger}(\{\varphi\}, t) 
i\hbar \frac{\partial}{\partial t}\widehat{\varPhi}(\{\varphi\}, t)
- \widehat{\varPhi}^{\dagger}(\{\varphi\}, t) \widehat{H}_{\varphi}
\widehat{\varPhi}(\{\varphi\}, t),
\label{L-varPhi}
\end{eqnarray}
where $\widehat{\varPhi}(\{\varphi\}, t)$ is a functional operator,
$\{\varphi\}$ and $t$ stand for a field of $\varphi$ and a time, respectively,
and $\widehat{H}_{\varphi}$ 
is the Hamiltonian operator in QFT containing $\varphi(\bm{x})$
and its functional derivatives $\delta/\delta \varphi(\bm{x})$ 
(see eq.~\eqref{H-varphi-S-der}).
For instance, $\widehat{H}_{\varphi}$ is given by
\begin{eqnarray}
\widehat{H}_{\varphi} 
= \int \frac{\delta}{\delta {\varphi}(\bm{x})} \left(- i \hbar c \bm{\alpha} \cdot \bm{\nabla}
+ \beta mc^2\right)
{\varphi}(\bm{x}) d^3x
\label{H-Dirac-QFT}
\end{eqnarray}
for a free Dirac fermion.

The canonical conjugate of $\widehat{\varPhi}(\{\varphi\}, t)$ is defined by
\begin{eqnarray}
\widehat{\varPi}(\{\varphi\}, t) \equiv 
\frac{\partial \widehat{{L}}_{\varPhi}}
{\partial (\partial\widehat{\varPhi}(\{\varphi\}, t)/\partial t)}
= i \hbar \widehat{\varPhi}^{\dagger}(\{\varphi\}, t),
\label{Pi}
\end{eqnarray}
and the Hamiltonian operator $\widehat{H}_{\varPhi}$ in QFFT is obtained as
\begin{align}
\widehat{H}_{\varPhi} 
&\equiv \widehat{\varPi}(\{\varphi\}, t) 
\frac{\partial\widehat{\varPhi}(\{\varphi\}, t)}{\partial t}
- \widehat{{L}}_{\varPhi}
\nonumber \\
&= \widehat{\varPhi}^{\dagger}(\{\varphi\}, t) \widehat{H}_{\varphi}
\widehat{\varPhi}(\{\varphi\}, t)
= \frac{1}{i\hbar} \widehat{\varPi}(\{\varphi\}, t) \widehat{H}_{\varphi}
\widehat{\varPhi}(\{\varphi\}, t).
\label{H-varPhi}
\end{align}
We notice that the Hamiltonian operator in QFFT is also constructed by sandwiching
the Hamiltonian operator in QFT between two functional operators,
and then it has the nesting structure.
We list explicit forms of $\widehat{H}_{\varPhi}$ 
for particles with spin 1/2 in appendix A.

We impose the following quantization conditions on functional operators\footnote{
If eigenvalues of $\widehat{H}_{\varphi}$ are bounded below,
this system can also be quantized using commutation relations
although the Lagrangian consists of a first time-derivative term
and a (sign-flipping of) Hamiltonian (see eq.~\eqref{L-varPhi}).
This feature is different from that of the quantization for particles with spin 1/2 in QFT.
In fact, when we impose commutation relations on Dirac or Weyl fermion,
the system becomes ill-defined with the advent of negative norm states.
It stems from the existence of negative-energy states in relativistic QM.},
\begin{eqnarray}
\left\{\widehat{\varPhi}(\{\varphi\}, t), \widehat{\varPi}(\{\varphi\}, t)\right\} 
= i \hbar,~~
\left\{\widehat{\varPhi}(\{\varphi\}, t), \widehat{\varPhi}(\{\varphi\}, t)\right\} = 0,~~
\left\{\widehat{\varPi}(\{\varphi\}, t), \widehat{\varPi}(\{\varphi\}, t)\right\} = 0
\label{anti-c-QFFT}
\end{eqnarray}
or
\begin{eqnarray}
\left[\widehat{\varPhi}(\{\varphi\}, t), \widehat{\varPi}(\{\varphi\}, t)\right] 
= i \hbar,~~
\left[\widehat{\varPhi}(\{\varphi\}, t), \widehat{\varPhi}(\{\varphi\}, t)\right] = 0,~~
\left[\widehat{\varPi}(\{\varphi\}, t), \widehat{\varPi}(\{\varphi\}, t)\right] = 0.
\label{c-QFFT}
\end{eqnarray}

Functional operators obey the Heisenberg's equation of motion:
\begin{eqnarray}
i\hbar \frac{\partial}{\partial t}\widehat{\varPhi}(\{\varphi\}, t)
= \left[\widehat{\varPhi}(\{\varphi\}, t), \widehat{H}_{\varPhi}\right],~~
i\hbar \frac{\partial}{\partial t}\widehat{\varPi}(\{\varphi\}, t) 
= \left[\widehat{\varPi}(\{\varphi\}, t), \widehat{H}_{\varPhi}\right].
\label{H-eqs-QFFT}
\end{eqnarray}
Using eqs.~\eqref{H-varPhi}, \eqref{anti-c-QFFT} and \eqref{H-eqs-QFFT},
we derive the equations:
\begin{eqnarray}
i\hbar \frac{\partial}{\partial t}\widehat{\varPhi}(\{\varphi\}, t) 
= \widehat{H}_{\varphi} \widehat{\varPhi}(\{\varphi\}, t),~~
i\hbar \frac{\partial}{\partial t}\widehat{\varPi}(\{\varphi\}, t) 
= -\widehat{\varPi}(\{\varphi\}, t) \widehat{H}_{\varphi},
\label{H-eqs2-QFFT}
\end{eqnarray}
and these equations agree with the Euler-Lagrange equation:
\begin{eqnarray}
\frac{d}{dt}\left(\frac{\partial \widehat{{L}}_{\varPhi}}
{\partial (\partial \widehat{\varPhi}^{\dagger}/\partial t)}\right) 
- \frac{\partial \widehat{{L}}_{\varPhi}}
{\partial \widehat{\varPhi}^{\dagger}}=0,~~
\frac{d}{dt}\left(\frac{\partial \widehat{{L}}_{\varPhi}}
{\partial (\partial \widehat{\varPhi}/\partial t)}\right) 
- \frac{\partial \widehat{{L}}_{\varPhi}}
{\partial \widehat{\varPhi}}=0,
\label{EL-eq-QFFT}
\end{eqnarray}
that derived from the action integral 
$\displaystyle{\widehat{S}_{\varPhi} 
= \int \widehat{{L}}_{\varPhi}dt}$,
based on a least action principle.

In the Schr\"{o}dinger picture, 
functional operators $\widehat{\varPhi}(\{\varphi\})$
and $\widehat{\varPi}(\{\varphi\})$ are independent of time, 
and they are related to those in the Heisenberg picture as
\begin{eqnarray}
\widehat{\varPhi}(\{\varphi\}, t) = e^{\frac{i}{\hbar}\widehat{H}_{\varPhi}t}
\widehat{\varPhi}(\{\varphi\}) e^{-\frac{i}{\hbar}\widehat{H}_{\varPhi}t},~~
\widehat{\varPi}(\{\varphi\}, t) = e^{\frac{i}{\hbar}\widehat{H}_{\varPhi}t}
\widehat{\varPi}(\{\varphi\}) e^{-\frac{i}{\hbar}\widehat{H}_{\varPhi}t}.
\label{varPhi-S}
\end{eqnarray}
Using eq.~\eqref{varPhi-S} and the conservation law of $\widehat{H}_{\varPhi}$, i.e.,
$\displaystyle{{d\widehat{H}_{\varPhi}}/{dt}=0}$, $\widehat{H}_{\varPhi}$ is rewritten 
in a time-independent form as
\begin{eqnarray}
\widehat{H}_{\varPhi} 
= \widehat{\varPhi}^{\dagger}(\{\varphi\}) 
\widehat{H}_{\varphi} \widehat{\varPhi}(\{\varphi\})
= \frac{1}{i\hbar} \widehat{\varPi}(\{\varphi\}) 
\widehat{H}_{\varphi} \widehat{\varPhi}(\{\varphi\}).
\label{H-varPhi-S}
\end{eqnarray}

Functional operators $\widehat{\varPhi}(\{\varphi\})$ and $\widehat{\varPi}(\{\varphi\})$
obey the anti-commutation relations:
\begin{eqnarray}
\left\{\widehat{\varPhi}(\{\varphi\}), \widehat{\varPi}(\{\varphi\})\right\} 
= i \hbar,~~
\left\{\widehat{\varPhi}(\{\varphi\}), \widehat{\varPhi}(\{\varphi\})\right\} = 0,~~
\left\{\widehat{\varPi}(\{\varphi\}), \widehat{\varPi}(\{\varphi\})\right\} = 0
\label{anti-c-QFFT-S}
\end{eqnarray}
or commutation relations:
\begin{eqnarray}
\left[\widehat{\varPhi}(\{\varphi\}), \widehat{\varPi}(\{\varphi\})\right] 
= i \hbar,~~
\left[\widehat{\varPhi}(\{\varphi\}), \widehat{\varPhi}(\{\varphi\})\right] = 0,~~
\left[\widehat{\varPi}(\{\varphi\}), \widehat{\varPi}(\{\varphi\})\right] = 0.
\label{c-QFFT-S}
\end{eqnarray}
From the first conditions in eqs.~\eqref{anti-c-QFFT-S} and \eqref{c-QFFT-S},
$\widehat{\varPi}(\{\varphi\})$ is given by
$\widehat{\varPi}(\{\varphi\}) 
= i\hbar {\delta}/{\delta {\varPhi}(\{\varphi\})}$
or $\widehat{\varPi}(\{\varphi\}) 
= -i\hbar {\delta}/{\delta {\varPhi}(\{\varphi\})}$, respectively,
in the representative diagonalizing $\widehat{\varPhi}(\{\varphi\})$
such as $\widehat{\varPhi}(\{\varphi\})|\varPhi\rangle 
= {\varPhi}(\{\varphi\})|\varPhi\rangle$
with a configuration of field's functional ${\varPhi}(\{\varphi\})$.

Any state $|\varPsi_{\rm M}(t)\rangle$ is evolved by the Schr\"{o}dinger equation:
\begin{eqnarray}
i \hbar \frac{d}{dt} |\varPsi_{\rm M}(t)\rangle = \widehat{H}_{\varPhi}
 |\varPsi_{\rm M}(t)\rangle,
\label{S-eq-QFFT}
\end{eqnarray}
and its formal solution is given by
\begin{eqnarray}
|\varPsi_{\rm M}(t)\rangle 
= e^{-\frac{i}{\hbar}\widehat{H}_{\varPhi}t} |\varPsi_{\rm M}(0)\rangle.
\label{S-eq-sol-QFFT}
\end{eqnarray}
Multiplying $\langle \varPhi|$ by the both sides of eq.~(\ref{S-eq-QFFT}),
we obtain the equation:
\begin{eqnarray}
i \hbar \frac{\partial}{\partial t} \varPsi_{\rm M}(\varPhi, t)
= \widehat{H}_{\varPhi} \varPsi_{\rm M}(\varPhi, t),
\label{S-eq-QFFT-functional}
\end{eqnarray}
where $\varPsi_{\rm M}(\varPhi, t)=\langle \varPhi|\varPsi_{\rm M}(t)\rangle$ 
is a state's functional in QFFT,
and $\widehat{H}_{\varPhi}$ is written by
\begin{eqnarray}
\widehat{H}_{\varPhi} 
= \frac{\delta}{\delta {\varPhi}(\{\varphi\})} \widehat{H}_{\varphi} {\varPhi}(\{\varphi\})
~~~{\rm or}~~~
\widehat{H}_{\varPhi} 
= - \frac{\delta}{\delta {\varPhi}(\{\varphi\})}\widehat{H}_{\varphi} {\varPhi}(\{\varphi\}),
\label{H-varphi-S-der-QFFT}
\end{eqnarray}
using $\widehat{\varPi}(\{\varphi\})
 = i\hbar {\delta}/{\delta {\varPhi}(\{\varphi\})}$
or $\widehat{\varPi}(\{\varphi\})
 = - i\hbar {\delta}/{\delta {\varPhi}(\{\varphi\})}$, respectively.

\subsection{Derivation of quantum field theory}

Let us investigate whether QFT can be rebuilt from QFFT or not.

Before proceeding to deal with it, we need to clarify a role of functional operators.
To get a hint, we list empirical laws concerning particles.
\begin{itemize}
\item Elementary particles create and annihilate.
Particles can appear in the vacuum, and a multi-particle state can return to the vacuum state.
The vacuum state plays a role as a basis to construct any physical states,
e.g., multi-particle states are obtained by multiplying creation operators.
\item There is a hierarchical structure on matters, e.g., 
atoms consist of an nucleus and electrons, 
nuclei are composed of nucleons (protons and neutrons), 
and nucleons are made up of quarks.
\item There is a hierarchical structure on physical laws, too.
Or equivalently, specific physical laws hold 
at each level of the structure on matters.
Physical laws are universal in our universe.
\end{itemize}

From the above laws, we form a conjecture
that {\it information on physical laws (particle contents and its related parameters)
is built in a vacuum state and the vacuum state is universal in our universe,
in a sense that it obeys common physical laws everywhere, 
although it can be differently described depending on a situation of observers
and an energy scale.
Our vacuum state keeps a potential to create some definite elementary particles
in accordance with physical systems.
And if our universe has a beginning, 
our vacuum state must be also created by an operation of anything.}

According to the conjecture, 
let us assume that $\widehat{\varPhi}^{\dagger}(\{\varphi\})$
is an operator to produce a vacuum state $|0\rangle_{\{\varphi\}}$
which has a potential to create an elementary particle $\varphi$
and $\varphi$ obeys definite laws of QFT.
This assumption is expressed by
\begin{eqnarray}
\widehat{\varPhi}^{\dagger}(\{\varphi\}) |\Slash\rangle = |0\rangle_{\{\varphi\}},
\label{Slash0}
\end{eqnarray}
where $|\Slash\rangle$ is ``nothingness state'' or ``empty state'', 
and it satisfies $\widehat{\varPhi}(\{\varphi\}) |\Slash\rangle = 0$
and $\langle\Slash |\Slash\rangle =1$.
Here, we list relations:
\begin{eqnarray}
\widehat{\varPhi}^{\dagger}(\{\varphi\}) |\Slash\rangle = |0\rangle_{\{\varphi\}},~~
\widehat{\varPhi}(\{\varphi\}) |\Slash\rangle = 0,~~
\langle\Slash |\widehat{\varPhi}(\{\varphi\}) =_{\{\varphi\}}\!\!\langle 0|,~~
\langle\Slash |\widehat{\varPhi}^{\dagger}(\{\varphi\}) = 0.
\label{Slash0s}
\end{eqnarray}
Note that $|\Slash\rangle$ is not a vacuum state in QFT
but a more fundamental one.
If $\widehat{\varPhi}^{\dagger}(\{\varphi\})$ produces a vacuum state,
it must be accompanied by an emergence of spacetime where $\varphi$ lives, 
and then an introduction of gravity is inevitable to formulate a complete theory.
We will discuss a topic that is pertinent to gravity in subsection 4.2. 

If $\widehat{\varPhi}^{\dagger}(\{\varphi\})$ 
and $\widehat{\varPhi}(\{\varphi\})$
play a role to install a vacuum state with a specific spacetime 
and to remove it, it would be suitable to
refer to $\widehat{\varPhi}^{\dagger}(\{\varphi\})$ and $\widehat{\varPhi}(\{\varphi\})$
as ``installation operator'' and ``removal operator'', respectively.
Then, ``Which came first particles or spacetime?'' on Q2 can be solved,
because particles and a spacetime can be installed 
at the same time by $\widehat{\varPhi}^{\dagger}(\{\varphi\})$ in QFFT.

Now let us derive QFT from QFFT, based on $\bm{x}$-representation in QM.

For $\widehat{\varPhi}^{\dagger}(\{\varphi\})$
satisfying the anti-commutation relations~\eqref{anti-c-QFFT-S},
a state $|\varPsi_{\rm M}(t)\rangle$ in QFFT is written as
\begin{eqnarray}
|\varPsi_{\rm M}(t)\rangle
= \varPsi^{(0)}(t) |\Slash\rangle
+ {\varPsi}(\{\varphi\},t)~\widehat{\varPhi}^{\dagger}(\{\varphi\})
|\Slash\rangle,
\label{stateM-fermion}
\end{eqnarray}
where ${\varPsi}(\{\varphi\},t)$ is a state's functional of one world
made of $\varphi$ and an expression of ${\varPsi}(\{\varphi\},t)$ 
will be given later (see eq.~\eqref{fuctional-M-1}).
In this case, one universe alone can appear and it has a literal meaning.

In contrast, for $\widehat{\varPhi}^{\dagger}(\{\varphi\})$
satisfying the commutation relations~\eqref{c-QFFT-S},
a state is written as a superposition of states 
constructed on various numbers of vacuum states,
\begin{eqnarray}
|\varPsi_{\rm M}(t)\rangle
= \varPsi^{(0)}(t) |\Slash\rangle
+ \sum_{M=1}^{\infty} {\varPsi}(\underbrace{\{\varphi\},\cdots,
\{\varphi\}}_{M},t)~
\frac{1}{\sqrt{M!}}
\left(\widehat{\varPhi}^{\dagger}(\{\varphi\})\right)^M
|\Slash\rangle,
\label{stateM-boson}
\end{eqnarray}
where ${\varPsi}(\underbrace{\{\varphi\},\cdots,\{\varphi\}}_{M},t)$
is a state's functional of $M$ worlds and 
$\frac{1}{\sqrt{M!}}
\left(\widehat{\varPhi}^{\dagger}(\{\varphi\})\right)^M|\Slash\rangle$
is a vacuum state described by
\begin{eqnarray}
\frac{1}{\sqrt{M!}}\left(\widehat{\varPhi}^{\dagger}(\{\varphi\})\right)^M |\Slash\rangle
= \underbrace{|0\rangle_{\{\varphi\}} \otimes \cdots \otimes |0\rangle_{\{\varphi\}}}_{M}.
\label{stateM-N}
\end{eqnarray}
In this case, $M$ worlds can be interpreted as $M$ identical universes
developing according to same physical laws.

Let us study a state on one world limited as
\begin{eqnarray}
|\varPsi_{\rm M}(t)\rangle^{(1)} = |\varPsi(t)\rangle 
= {\varPsi}(\{\varphi\},t)~\widehat{\varPhi}^{\dagger}(\{\varphi\})
|\Slash\rangle
= {\varPsi}(\{\varphi\},t)|0\rangle,
\label{stateM-1}
\end{eqnarray}
where ${\varPsi}(\{\varphi\},t)$ is a state's functional and it is depicted by
\begin{eqnarray}
{\varPsi}(\{\varphi\},t)
= \psi^{(0)}(t)+
\sum_{N = 1}^{\infty} \int d^3x_1 \cdots d^3x_N~
\psi(\bm{x}_1, \cdots, \bm{x}_N, t) \frac{1}{\sqrt{N!}}
\varphi^{\dagger}(\bm{x}_1) \cdots \varphi^{\dagger}(\bm{x}_N),
\label{fuctional-M-1}
\end{eqnarray}
using eq.~\eqref{B-fuctional-M-1}.
Note that ${\varPsi}(\{\varphi\},t)$ should not be confused with the wave functional  
$\varPsi(\varphi, t) \equiv \langle\varphi |\varPsi(t)\rangle$ in QFT.
As seen in eq.~\eqref{B-fuctional-M-1-QFT}, they are related to each other as
\begin{eqnarray}
\varPsi(\varphi, t) = {\varPsi}(\{\varphi\},t) \mathscr{V}_0(\varphi),
\label{fuctional-M-1-QFT}
\end{eqnarray}
with $\mathscr{V}_0(\varphi) = \langle\varphi|0\rangle$.
Here and hereafter, we omit a subscript ${\{\varphi\}}$
attached the vacuum state $|0\rangle$ and $\langle 0|$ 
in this subsection, for simplicity.

The vacuum wave function is defined by
\begin{align}
{\varPsi}_{\rm V}(t) &\equiv \langle 0|\varPsi_{\rm M}(t)\rangle
= \langle\Slash |\widehat{\varPhi}(\{\varphi\})|\varPsi_{\rm M}(t)\rangle
\nonumber \\
&= \langle\Slash |e^{\frac{i}{\hbar}\widehat{H}_{\varPhi}t}
\widehat{\varPhi}(\{\varphi\})e^{-\frac{i}{\hbar}\widehat{H}_{\varPhi}t}
|\varPsi_{\rm M}(0)\rangle
= \langle\Slash |\widehat{\varPhi}(\{\varphi\},t)|\varPsi_{\rm M}(0)\rangle,
\label{varPsiV}
\end{align}
using eq.~\eqref{S-eq-sol-QFFT},
$\langle\Slash|e^{\frac{i}{\hbar}\widehat{H}_{\varPhi}t}=\langle\Slash|$, i.e.,
$\langle\Slash|\widehat{H}_{\varPhi} = 0$ and eq.~\eqref{varPhi-S}.
From eq.~\eqref{varPsiV} and the first equation in eq.~\eqref{H-eqs2-QFFT},
we find that ${\varPsi}_{\rm V}(t)$ and $\widehat{\varPhi}(\{\varphi\},t)$ satisfy
the same type of equation:
\begin{eqnarray}
i\hbar \frac{\partial}{\partial t}{\varPsi}_{\rm V}(t) = \mathscr{E}_0 {\varPsi}_{\rm V}(t),~~
i\hbar \frac{\partial}{\partial t}\widehat{\varPhi}(\{\varphi\},t)
 = \mathscr{E}_0 \widehat{\varPhi}(\{\varphi\},t),
\label{varPsiV-eq}
\end{eqnarray}
where $\mathscr{E}_0 = \langle 0|\widehat{H}_{\varphi}|0\rangle$
and $\widehat{H}_{\varphi}$ is the Hamiltonian operator in QFT.
Then, the time evolution of
$\widehat{\varPhi}(\{\varphi\},t)$ and $\widehat{\varPhi}^{\dagger}(\{\varphi\},t)$
is determined as
\begin{eqnarray}
\widehat{\varPhi}(\{\varphi\},t) = e^{-\frac{i}{\hbar}\mathscr{E}_0 t}
\widehat{\varPhi}(\{\varphi\}),~~
\widehat{\varPhi}^{\dagger}(\{\varphi\},t) = \widehat{\varPhi}^{\dagger}(\{\varphi\})
e^{\frac{i}{\hbar}\mathscr{E}_0 t}.
\label{functional-op-sol}
\end{eqnarray}
Note that ${\varPsi}_{\rm V}(t)$ agrees with
$\langle 0|\varPsi(t)\rangle=\psi^{(0)}(t)$ in QFT
and it satisfies eq.~\eqref{B-varPsiV}.

From eqs.~\eqref{S-eq-QFFT} and \eqref{stateM-1},
we can derive the Schr\"{o}dinger equation in QFT:
\begin{eqnarray}
i \hbar \frac{d}{dt} |\varPsi(t)\rangle
= \widehat{H}_{\varphi}|\varPsi(t)\rangle,
\label{S-eq-QFT-state}
\end{eqnarray}
under the assumption that $\widehat{H}_{\varphi}$ is the Hamiltonian operator in QFT, 
as follows,
\begin{align}
i \hbar \frac{d}{dt} |\varPsi(t)\rangle
&= i \hbar \frac{d}{dt} |\varPsi_{\rm M}(t)\rangle^{(1)}
= \widehat{H}_{\varPhi}|\varPsi_{\rm M}(t)\rangle^{(1)}
= \widehat{\varPhi}^{\dagger}(\{\varphi\}) 
\widehat{H}_{\varphi} \widehat{\varPhi}(\{\varphi\})|\varPsi(t)\rangle
\nonumber \\
&= \widehat{\varPhi}^{\dagger}(\{\varphi\}) 
\widehat{H}_{\varphi} \widehat{\varPhi}(\{\varphi\})~
{\varPsi}(\{\varphi\},t)~\widehat{\varPhi}^{\dagger}(\{\varphi\})
|\Slash\rangle
= \widehat{H}_{\varphi}|\varPsi(t)\rangle.
\label{S-eq-QFT-state-der}
\end{align}
Multiplying $\langle \varphi|$ by both sides of eq.(\ref{S-eq-QFT-state}),
we obtain the equation on the wave functional in QFT:
\begin{eqnarray}
i \hbar \frac{\partial}{\partial t} \varPsi(\varphi, t)
= \widehat{H}_{\varphi} \varPsi(\varphi, t).
\label{S-eq-QFT-functional}
\end{eqnarray}

Next let us justify that a fermion (boson) field $\varphi(\bm{x})$
is quantized in order to satisfy the anti-commutation relations 
(the commutation relations).
Under an infinitesimal translation $\bm{x}' = \bm{x} - \bm{\varepsilon}$,
the field $\varphi(\bm{x})$ transforms as
$\varphi'(\bm{x}') = \varphi'(\bm{x} - \bm{\varepsilon}) =\varphi(\bm{x})$,
and then an infinitesimal change of $\varphi(\bm{x})$ is given by
\begin{eqnarray}
\delta_{\varepsilon} \varphi(\bm{x}) = \varphi'(\bm{x}) - \varphi(\bm{x})
= \varphi(\bm{x}+\bm{\varepsilon}) - \varphi(\bm{x})
= \varepsilon^i \partial_i \varphi(\bm{x}).
\label{delta-varphi-cl}
\end{eqnarray}
If its field's functional $\widehat{\varPhi}(\{\varphi\}, t)$ transforms as
\begin{eqnarray}
\widehat{\varPhi}'(\{\varphi'\}, t) = 
\widehat{\varPhi}'(\{\varphi + \delta_{\varepsilon}\varphi\}, t) 
= \widehat{\varPhi}(\{\varphi\}, t)
\label{varphi'}
\end{eqnarray}
under $\bm{x}' = \bm{x} - \bm{\varepsilon}$,
$\widehat{\varPhi}(\{\varphi\}, t)$ changes infinitesimally as
\begin{align}
\delta_{\varepsilon} \widehat{\varPhi}(\{\varphi\}, t) 
&= \widehat{\varPhi}'(\{\varphi\}, t) - \widehat{\varPhi}(\{\varphi\}, t)
= \widehat{\varPhi}(\{\varphi-\delta_{\varepsilon}\varphi\}, t) 
- \widehat{\varPhi}(\{\varphi\}, t)
\nonumber \\
&= - \left(\widehat{\varPhi}(\{\varphi\}, t) 
- \widehat{\varPhi}(\{\varphi-\delta_{\varepsilon}\varphi\}, t)\right)
= - \int  d^3x~\delta_{\varepsilon} \varphi(\bm{x}) \frac{\delta}{\delta \varphi(\bm{x})}
\widehat{\varPhi}(\{\varphi\}, t)
\nonumber \\
&= - \int  d^3x~\varepsilon^i \partial_i \varphi(\bm{x}) \frac{\delta}{\delta \varphi(\bm{x})}
\widehat{\varPhi}(\{\varphi\}, t).
\label{delta-varPhi}
\end{align}
Note that eq.~\eqref{varphi'} implies the translational invariance of the vacuum state,
which is one of features in relativistic QFT.

Then, the action integral 
$\displaystyle{\widehat{S}_{\varPhi} = \int \widehat{L}_{\varPhi} dt}$ changes as
\begin{align}
\delta_{\varepsilon} \widehat{S}_{\varPhi} 
&= \int \delta_{\varepsilon} \widehat{L}_{\varPhi} dt
= \int \frac{\partial}{\partial t} \left(\widehat{\varPhi}^{\dagger}(\{\varphi\}, t)
i\hbar \delta_{\varepsilon} \widehat{\varPhi}(\{\varphi\}, t)\right)dt
\nonumber \\
&= \int \frac{\partial}{\partial t} \left\{\widehat{\varPhi}^{\dagger}(\{\varphi\}, t)
\int d^3x~\varepsilon^i \partial_i \varphi(\bm{x}) 
\left(-i\hbar \frac{\delta}{\delta \varphi(\bm{x})}\right)
\widehat{\varPhi}(\{\varphi\}, t)\right\}dt,
\label{delta-SvarPhi}
\end{align}
under the condition that $\widehat{\varPhi}(\{\varphi\}, t)$ obeys 
the equation of motion \eqref{H-eqs2-QFFT}.
From eq.~\eqref{delta-SvarPhi}, the momentum operator $\widehat{P}_{\varPhi~i}$ 
can be read off as
\begin{align}
\widehat{P}_{\varPhi~i} &\equiv
\widehat{\varPhi}^{\dagger}(\{\varphi\}, t)
\int d^3x~\partial_i \varphi(\bm{x}) 
\left(-i\hbar \frac{\delta}{\delta \varphi(\bm{x})}\right)
\widehat{\varPhi}(\{\varphi\}, t)
\nonumber \\
&= \widehat{\varPhi}^{\dagger}(\{\varphi\}, t)
\left(\int d^3x~\widehat{\pi}(\bm{x}) \partial_i \widehat{\varphi}(\bm{x}) \right)
\widehat{\varPhi}(\{\varphi\}, t)
\nonumber \\
&= \widehat{\varPhi}^{\dagger}(\{\varphi\}, t)
\left(\frac{1}{c} \int d^3x~{\widehat{T}^0}_{~i}(\bm{x})\right)
\widehat{\varPhi}(\{\varphi\}, t),
\label{delta-PvarPhi}
\end{align}
where we use $\widehat{\varphi}(\bm{x}) = {\varphi}(\bm{x})$
and $\widehat{\pi}(\bm{x}) = i\hbar{\delta}/{\delta \varphi(\bm{x})}$
for fermion
($\widehat{\varphi}(\bm{x}) = {\varphi}(\bm{x})$
and $\widehat{\pi}(\bm{x}) = -i\hbar {\delta}/{\delta \varphi(\bm{x})}$
for boson), and ${\widehat{T}}^0_{~i}(\bm{x})$ 
is a component of the energy-momentum tensor in QFT.
Note that $\varphi(\bm{x})$ is a Grassmann variable for fermion
and a constant term in the integration is subtracted.

In this way, we find that there also exists a nested structure for momenta
and verify that fields become operators 
and satisfy the anti-commutation relations for fermion (the commutation relations for boson).
We notice that these features come from the transformation property of field's functional
and an answer to the question 
``Why are particles or fields quantized in the first place?'' is obtained.

Let us evaluate the expectation value of momenta for the state
given by eq.~\eqref{stateM-1}.
It is calculated as
\begin{align}
&^{(1)}\langle \varPsi_{\rm M}(t)|\widehat{\bm{P}}_{\varPhi}|\varPsi_{\rm M}(t)\rangle^{(1)}
= \langle \varPsi(t)|\widehat{\bm{P}}_{\varPhi}|\varPsi(t)\rangle
\nonumber \\
&~~~~= \langle \Slash |\widehat{\varPhi}(\{\varphi\}){\varPsi}^{\dagger}(\{\varphi\},t)
\widehat{\varPhi}^{\dagger}(\{\varphi\})
\left(\int d^3x~\widehat{\pi}(\bm{x}) \bm{\nabla} \widehat{\varphi}(\bm{x}) \right)
\widehat{\varPhi}(\{\varphi\})
{\varPsi}(\{\varphi\},t)\widehat{\varPhi}^{\dagger}(\{\varphi\})|\Slash\rangle
\nonumber \\
&~~~~= {\varPsi}^{\dagger}(\{\varphi\},t)
\left(\int d^3x~\widehat{\pi}(\bm{x}) \bm{\nabla} \widehat{\varphi}(\bm{x}) \right)
{\varPsi}(\{\varphi\},t)
= \int \mathscr{D}\varphi~ {\varPsi}^{\dagger}(\varphi,t)\widehat{\bm{P}}{\varPsi}(\varphi,t)
\nonumber \\
&~~~~= \langle \varPsi(t)|\widehat{\bm{P}}|\varPsi(t)\rangle,
\label{<P>-QFFT}
\end{align}
using the conditions \eqref{anti-c-QFFT-S} or \eqref{c-QFFT-S}, 
$\langle \Slash | \Slash \rangle = 1$, and eq.~\eqref{B-<Omega>}.

In the same way, the expectation value of energy for the state
given by eq.~\eqref{stateM-1} is calculated as
\begin{align}
&^{(1)}\langle \varPsi_{\rm M}(t)|\widehat{H}_{\varPhi}|\varPsi_{\rm M}(t)\rangle^{(1)}
= \langle \varPsi(t)|\widehat{H}_{\varPhi}|\varPsi(t)\rangle
= {\varPsi}^{\dagger}(\{\varphi\},t)\widehat{H}_{\varphi}
{\varPsi}(\{\varphi\},t)
\nonumber \\
&~~~~
= \int \mathscr{D}\varphi~ {\varPsi}^{\dagger}(\varphi,t)\widehat{H}_{\varphi}{\varPsi}(\varphi,t)
= \langle \varPsi(t)|\widehat{H}_{\varphi}|\varPsi(t)\rangle.
\label{<H>-QFFT}
\end{align}

In this way, we arrive at the expression for expectation values in QFT,
and the expectation values in QM are also obtained by fixing a number of particles,
as seen in subsection 2.2.

\section{Universes with different particle contents}

Now it is time to tackle Q3, i.e., 
``Why is our universe described by the standard model at the electroweak scale?''.
Under the precondition that the existence of our universe can be
understood by the combination of the level II multiverse 
(an ensemble of foreign universes) and the anthropic principle,
and the above question is not really acknowledged as a problem,
we face the question whether a framework 
to describe the level II multiverse can be constructed or not.
In the following, we investigate a theoretical framework to deal with 
a set of universes with different elementary particles and parameters
based on the physical laws of QFT.

\subsection{Level II multiverse}

We extend our formulation to 
an assembly of universes with different elementary particles and parameters.

Let us first introduce installation operators 
$\displaystyle{\widehat{\varPhi}_{(a)}^{\dagger}\left(\{\varphi_{k^{(a)}}\}\right)}$ 
which produce a vacuum state $|0\rangle_{\{\varphi_{k^{(a)}}\}}$
where definite elementary particles $\varphi_{k^{(a)}}$ 
can be created and work obeying the laws of QFT.
This is expressed by
\begin{eqnarray}
\widehat{\varPhi}_{(a)}^{\dagger}\left(\{\varphi_{k^{(a)}}\}\right) |\Slash\rangle 
= |0\rangle_{\{\varphi_{k^{(a)}}\}},
\label{Slash0-k}
\end{eqnarray}
where $|\Slash\rangle$ is the nothingness state and it satisfies
$\widehat{\varPhi}_{(a)}(\{\varphi_{k^{(a)}}\}) |\Slash\rangle = 0$
and $\langle\Slash |\Slash\rangle =1$.
Here, $a(=1, \cdots, \mathscr{N})$ is a label which specifies universes
with a set of particles $\{\varphi_{k^{(a)}}\}$ and parameters, 
and $k^{(a)}$ is a label which represents particles
\footnote{
For the sake of completeness, installation operators and vacuum states
should be denoted as 
$\displaystyle{\widehat{\varPhi}_{(a)}^{\dagger}\left(\{\varphi_{k^{(a)}}\},
\{\alpha_{j^{(a)}}\}\right)}$ and $|0\rangle_{\{\varphi_{k^{(a)}}\},\{\alpha_{j^{(a)}}\}}$,
respectively, where $\{\alpha_{j^{(a)}}\}$ is a label which represents
a set of parameters,
but here and hereafter $\{\alpha_{j^{(a)}}\}$ is omitted to avoid complications.}.
We have relations such as
\begin{eqnarray}
&~& \widehat{\varPhi}_{(a)}^{\dagger}\left(\{\varphi_{k^{(a)}}\}\right)
|\Slash\rangle = |0\rangle_{\{\varphi_{k^{(a)}}\}},~~
\widehat{\varPhi}_{(a)}\left(\{\varphi_{k^{(a)}}\}\right) |\Slash\rangle = 0,~~
\nonumber \\
&~& \langle\Slash |\widehat{\varPhi}_{(a)}\left(\{\varphi_{k^{(a)}}\}\right) 
=_{\{\varphi_{k^{(a)}}\}}\!\!\langle 0|,~~
\langle\Slash |\widehat{\varPhi}_{(a)}^{\dagger}\left(\{\varphi_{k^{(a)}}\}\right) = 0,
\label{Slash0s-k}
\end{eqnarray}
and impose the following conditions on 
$\widehat{\varPhi}_{(a)}\left(\{\varphi_{k^{(a)}}\}\right)$
and $\widehat{\varPhi}_{(a)}^{\dagger}\left(\{\varphi_{k^{(a)}}\}\right)$:
\begin{eqnarray}
&~& \left\{\widehat{\varPhi}_{(a)}\left(\{\varphi_{k^{(a)}}\}\right), 
\widehat{\varPhi}_{(b)}^{\dagger}\left(\{\varphi_{k^{(b)}}\}\right)\right\} 
= \delta_{ab},~~
\left\{\widehat{\varPhi}_{(a)}\left(\{\varphi_{k^{(a)}}\}\right), 
\widehat{\varPhi}_{(b)}\left(\{\varphi_{k^{(b)}}\}\right)\right\} = 0,~~
\nonumber \\
&~& \left\{\widehat{\varPhi}_{(a)}^{\dagger}\left(\{\varphi_{k^{(a)}}\}\right), 
\widehat{\varPhi}_{(b)}^{\dagger}\left(\{\varphi_{k^{(b)}}\}\right)\right\} = 0
\label{anti-c-QFFT-k}
\end{eqnarray}
or
\begin{eqnarray}
&~& \left[\widehat{\varPhi}_{(a)}\left(\{\varphi_{k^{(a)}}\}\right), 
\widehat{\varPhi}_{(b)}^{\dagger}\left(\{\varphi_{k^{(b)}}\}\right)\right]
= \delta_{ab},~~
\left[\widehat{\varPhi}_{(a)}\left(\{\varphi_{k^{(a)}}\}\right), 
\widehat{\varPhi}_{(b)}\left(\{\varphi_{k^{(b)}}\}\right)\right] = 0,~~
\nonumber \\
&~& \left[\widehat{\varPhi}_{(a)}^{\dagger}\left(\{\varphi_{k^{(a)}}\}\right), 
\widehat{\varPhi}_{(b)}^{\dagger}\left(\{\varphi_{k^{(b)}}\}\right)\right] = 0.
\label{c-QFFT-k}
\end{eqnarray}

Because a production of the vacuum state $|0\rangle_{\{\varphi_{k^{(a)}}\}}$ by
$\displaystyle{\widehat{\varPhi}_{(a)}^{\dagger}\left(\{\varphi_{k^{(a)}}\}\right)}$ 
must be associated with the emergence of spacetime,
a graviton appears inevitably and each universe
has the invariance under the general transformation of coordinates.
Its consequence will be discussed in the next subsection.

When field's functionals satisfy eq.~\eqref{anti-c-QFFT-k},
a state $|\varPsi_{\rm M_{II}}(t)\rangle$ in the level II multiverse is written as
a superposition of states constructed on various numbers of vacuum states as follows,
\begin{align}
|\varPsi_{\rm M_{II}}(t)\rangle
&= \varPsi^{(0)}(t) |\Slash\rangle
+ \sum_{M=1}^{\mathscr{N}} \sum_{a_1=1}^{\mathscr{N}} \cdots \sum_{a_M=1}^{\mathscr{N}}
{\varPsi}\left(\{\varphi_{k^{(a_1)}}\},\cdots,
\{\varphi_{k^{(a_M)}}\},t\right)~
\nonumber \\
&~ ~~~~~~~~~~ 
\times \widehat{\varPhi}_{(a_1)}^{\dagger}\left(\{\varphi_{k^{(a_1)}}\}\right)
\cdots \widehat{\varPhi}_{(a_M)}^{\dagger}\left(\{\varphi_{k^{(a_M)}}\}\right)
|\Slash\rangle,
\label{stateM-anti-c-k}
\end{align}
where the summation is taken in the range $a_1 > \cdots > a_M$,\footnote{
When a label $a_n$ takes continuous values, 
a summation should be replaced into an integration.
Same applies to the following.}
$t$ is an auxiliary parameter which is identified as a time
when a system is limited in some observable region of our universe,
${\varPsi}\left(\{\varphi_{k^{(a_1)}}\},\cdots,
\{\varphi_{k^{(a_M)}}\},t\right)$
is a state's functional related to $M$ kinds of universes, 
and $\widehat{\varPhi}_{(a_1)}^{\dagger}\left(\{\varphi_{k^{(a_1)}}\}\right)
\cdots \widehat{\varPhi}_{(a_M)}^{\dagger}\left(\{\varphi_{k^{(a_M)}}\}\right)
|\Slash\rangle$ is a vacuum state described by
\begin{eqnarray}
\widehat{\varPhi}_{(a_1)}^{\dagger}\left(\{\varphi_{k^{(a_1)}}\}\right)
\cdots \widehat{\varPhi}_{(a_M)}^{\dagger}\left(\{\varphi_{k^{(a_M)}}\}\right)
|\Slash\rangle
= |0\rangle_{\{\varphi_{k^{(a_1)}}\}} \otimes 
\cdots \otimes |0\rangle_{\{\varphi_{k^{(a_M)}}\}}.
\label{stateM-N-k}
\end{eqnarray}

When field's functionals satisfy eq.~\eqref{c-QFFT-k},
$|\varPsi_{\rm M_{II}}(t)\rangle$ is written by
\begin{align}
|\varPsi_{\rm M_{II}}(t)\rangle
&= \varPsi^{(0)}(t) |\Slash\rangle 
\nonumber \\
&~~~~~~~~~~~ 
+ \sum_{M_{(a_1)}=1}^{\infty} \cdots \sum_{M_{(a_{\mathscr{N}})=1}}^{\infty}
{\varPsi}(\underbrace{\{\varphi_{k^{(a_1)}}\},\cdots,\{\varphi_{k^{(a_1)}}\}}_{M_{(a_1)}}
,\cdots,
\underbrace{\{\varphi_{k^{(a_{\mathscr{N}})}}\}, \cdots, 
\{\varphi_{k^{(a_{\mathscr{N}})}}\}}_{M_{(a_{\mathscr{N}})}}, t)~
\nonumber \\
&~~~~~~~~~~~ \times \frac{1}{\sqrt{M_{(a_1)}!}}\left(\widehat{\varPhi}_{(a_1)}^{\dagger}
\left(\{\varphi_{k^{(a_1)}}\}\right)\right)^{M_{(a_1)}}
\cdots \frac{1}{\sqrt{M_{(a_{\mathscr{N}})}!}}
\left(\widehat{\varPhi}_{(a_{\mathscr{N}})}^{\dagger}
\left(\{\varphi_{k^{(a_{\mathscr{N}})}}\}\right)\right)^{M_{(a_{\mathscr{N}})}}
|\Slash\rangle.
\label{stateM-c-k}
\end{align}

Any state $|\varPsi_{\rm M_{II}}(t)\rangle$ is evolved by the Schr\"{o}dinger equation:
\begin{eqnarray}
i \hbar \frac{d}{dt} |\varPsi_{\rm M_{II}}(t)\rangle 
= \widehat{H}_{\{\varPhi\}} |\varPsi_{\rm M_{II}}(t)\rangle,
\label{S-eq-QFFT-k}
\end{eqnarray}
and its formal solution is given by
\begin{eqnarray}
|\varPsi_{\rm M_{II}}(t)\rangle 
= e^{-\frac{i}{\hbar}\widehat{H}_{\{\varPhi\}}t} |\varPsi_{\rm M_{II}}(0)\rangle,
\label{S-eq-sol-QFFT-k}
\end{eqnarray}
where $\widehat{H}_{\{\varPhi\}}$ is the Hamiltonian operator in QFFT given by
\begin{eqnarray}
\widehat{H}_{\{\varPhi\}}
= \sum_{a=1}^{\mathscr{N}}
\widehat{\varPhi}^{\dagger}_{(a)}\left(\{\varphi_{k^{(a)}}\}\right) 
\widehat{H}_{\{\varphi_{k^{(a)}}\}}
\widehat{\varPhi}_{(a)}\left(\{\varphi_{k^{(a)}}\}\right).
\label{H-varPhi-k-S}
\end{eqnarray}
In eq.~\eqref{H-varPhi-k-S}, $\widehat{H}_{\{\varphi_{k^{(a)}}\}}$ 
are Hamiltonian operators in QFT containing $\varphi_{k^{(a)}}(x)$
and its functional derivatives $\delta/\delta \varphi_{k^{(a)}}(x)$,
and we see that the nested construction is realized.

In the Heisenberg picture, installation and removal operators
are given in a form with the time-dependence by
\begin{eqnarray}
&~& \widehat{\varPhi}_{(a)}^{\dagger}\left(\{\varphi_{k^{(a)}}\}, t\right) 
= e^{\frac{i}{\hbar}\widehat{H}_{\{\varPhi\}}t}
\widehat{\varPhi}_{(a)}^{\dagger}\left(\{\varphi_{k^{(a)}}\}\right) 
e^{-\frac{i}{\hbar}\widehat{H}_{\{\varPhi\}}t},~~
\nonumber \\
&~& \widehat{\varPhi}_{(a)}\left(\{\varphi_{k^{(a)}}\}, t\right) 
= e^{\frac{i}{\hbar}\widehat{H}_{\{\varPhi\}}t}
\widehat{\varPhi}_{(a)}\left(\{\varphi_{k^{(a)}}\}\right) 
e^{-\frac{i}{\hbar}\widehat{H}_{\{\varPhi\}}t},
\label{varPhi-HS-k}
\end{eqnarray}
and they are evolved by the Heisenberg's equation of motion:
\begin{eqnarray}
&~& i\hbar \frac{\partial}{\partial t}
\widehat{\varPhi}_{(a)}^{\dagger}\left(\{\varphi_{k^{(a)}}\}, t\right)
= \left[\widehat{\varPhi}_{(a)}^{\dagger}\left(\{\varphi_{k^{(a)}}\}, t\right), 
\widehat{H}_{\{\varPhi\}}\right],~~
\nonumber \\
&~& i\hbar \frac{\partial}{\partial t}
\widehat{\varPhi}_{(a)}\left(\{\varphi_{k^{(a)}}\}, t\right)
= \left[\widehat{\varPhi}_{(a)}\left(\{\varphi_{k^{(a)}}\}, t\right), 
\widehat{H}_{\{\varPhi\}}\right].
\label{H-eqs-QFFT-k}
\end{eqnarray}

The dynamics is summarized by the action integral:
\begin{align}
\widehat{S}_{\{\varPhi\}} &= \int \widehat{L}_{\{\varPhi\}} dt
\nonumber \\
&= \int \sum_{a=1}^{\mathscr{N}}
\left(\widehat{\varPhi}_{(a)}^{\dagger}\left(\{\varphi_{k^{(a)}}\}, t\right)
i\hbar \frac{\partial}{\partial t}
\widehat{\varPhi}_{(a)}\left(\{\varphi_{k^{(a)}}\}, t\right) \right.
\nonumber \\
&~ ~~~~~~~~~~~ \left.
- \widehat{\varPhi}_{(a)}^{\dagger}\left(\{\varphi_{k^{(a)}}\}, t\right) 
\widehat{H}_{\{\varphi_{k^{(a)}}\}}
\widehat{\varPhi}_{(a)}\left(\{\varphi_{k^{(a)}}\}, t\right)\right)dt.
\label{S-varPhi-k}
\end{align}

The canonical conjugate of $\widehat{\varPhi}_{(a)}\left(\{\varphi_{k^{(a)}}\}, t\right)$ 
is defined by
\begin{eqnarray}
\widehat{\varPi}_{(a)}\left(\{\varphi_{k^{(a)}}\}, t\right) \equiv 
\frac{\partial \widehat{{L}}_{\{\varPhi\}}}
{\partial (\partial\widehat{\varPhi}_{(a)}\left(\{\varphi_{k^{(a)}}\}, t\right)/\partial t)}
= i \hbar \widehat{\varPhi}^{\dagger}_{(a)}\left(\{\varphi_{k^{(a)}}\}, t\right),
\label{Pi-k}
\end{eqnarray}
and the Hamiltonian operator $\widehat{H}_{\{\varPhi\}}$ in QFFT is obtained as
\begin{align}
\widehat{H}_{\{\varPhi\}} 
&\equiv \sum_{a=1}^{\mathscr{N}}\widehat{\varPi}_{(a)}\left(\{\varphi_{k^{(a)}}\}, t\right)
\frac{\partial\widehat{\varPhi}_{(a)}\left(\{\varphi_{k^{(a)}}\}, t\right)}{\partial t}
- \widehat{{L}}_{\{\varPhi\}}
\nonumber \\
&= \sum_{a=1}^{\mathscr{N}}
\widehat{\varPhi}^{\dagger}_{(a)}\left(\{\varphi_{k^{(a)}}\}, t\right) 
\widehat{H}_{\{\varphi_{k^{(a)}}\}}
\widehat{\varPhi}_{(a)}\left(\{\varphi_{k^{(a)}}\}, t\right)
\nonumber \\
&= \frac{1}{i\hbar}\sum_{a=1}^{\mathscr{N}}
 \widehat{\varPi}_{(a)}\left(\{\varphi_{k^{(a)}}\}, t\right)
 \widehat{H}_{\{\varphi_{k^{(a)}}\}}
\widehat{\varPhi}_{(a)}\left(\{\varphi_{k^{(a)}}\}, t\right).
\label{H-varPhi-k}
\end{align}
It is shown that this $\widehat{H}_{\{\varPhi\}}$ 
agrees with that in eq.~\eqref{H-varPhi-k-S}, using the solutions
of functional operators (see eq.~\eqref{functional-op-sol-k}).

The following quantization conditions are imposed on field's functionals,
\begin{eqnarray}
&~& \left\{\widehat{\varPhi}_{(a)}\left(\{\varphi_{k^{(a)}}\}, t\right), 
\widehat{\varPi}_{(b)}\left(\{\varphi_{k^{(b)}}\}, t\right)\right\} 
= i \hbar\delta_{ab},~~
\left\{\widehat{\varPhi}_{(a)}\left(\{\varphi_{k^{(a)}}\}, t\right), 
\widehat{\varPhi}_{(b)}\left(\{\varphi_{k^{(b)}}\}, t\right)\right\} = 0,~~
\nonumber \\
&~& \left\{\widehat{\varPi}_{(a)}\left(\{\varphi_{k^{(a)}}\}, t\right), 
\widehat{\varPi}_{(b)}\left(\{\varphi_{k^{(b)}}\}, t\right)\right\} = 0
\label{anti-c-QFFT-t-k}
\end{eqnarray}
or
\begin{eqnarray}
&~& \left[\widehat{\varPhi}_{(a)}\left(\{\varphi_{k^{(a)}}\}, t\right), 
\widehat{\varPi}_{(b)}\left(\{\varphi_{k^{(b)}}\}, t\right)\right] 
= i \hbar\delta_{ab},~~
\left[\widehat{\varPhi}_{(a)}\left(\{\varphi_{k^{(a)}}\}, t\right), 
\widehat{\varPhi}_{(b)}\left(\{\varphi_{k^{(b)}}\}, t\right)\right] = 0,~~
\nonumber \\
&~& \left[\widehat{\varPi}_{(a)}\left(\{\varphi_{k^{(a)}}\}, t\right), 
\widehat{\varPi}_{(b)}\left(\{\varphi_{k^{(b)}}\}, t\right)\right] = 0,
\label{c-QFFT-t-k}
\end{eqnarray}
and then the Heisenberg's equation of motion \eqref{H-eqs-QFFT-k}
agrees with the Euler-Lagrange equation derived from the action integral 
\eqref{S-varPhi-k}, based on a least action principle.

\subsection{Physical implications}

Under the assumption that each universe contains specific elementary particle contents
including a graviton and physical parameters, and it is evolved by physical laws of QFT,
we discuss physical implications on a beginning of the universe based on QFFT.

\subsubsection{Inflation}

As a reference of the discussion from eqs.~\eqref{varPsiV}  to \eqref{functional-op-sol},
we obtain the solutions:
\begin{eqnarray}
\widehat{\varPhi}_{(a)}\left(\{\varphi_{k^{(a)}}\},t\right) 
= e^{-\frac{i}{\hbar}\mathscr{E}_{0}^{(a)} t}
\widehat{\varPhi}_{(a)}\left(\{\varphi_{k^{(a)}}\}\right),~~
\widehat{\varPhi}^{\dagger}_{(a)}\left(\{\varphi_{k^{(a)}}\},t\right) 
= \widehat{\varPhi}^{\dagger}_{(a)}\left(\{\varphi_{k^{(a)}}\}\right)
e^{\frac{i}{\hbar}\mathscr{E}_{0}^{(a)} t},
\label{functional-op-sol-k}
\end{eqnarray}
where $\mathscr{E}_{0}^{(a)} = \langle 0|\widehat{H}_{\{\varphi_{k^{(a)}}\}}|0\rangle$
is the vacuum energy of the universe labeled by $a$.
And the vacuum wave function $\varPsi_{\rm V}^{(a)}(t)$ is evolved as
\begin{eqnarray}
\varPsi_{\rm V}^{(a)}(t) = e^{-\frac{i}{\hbar}\mathscr{E}_{0}^{(a)} t}\varPsi_{\rm V}^{(a)}(0).
\label{vac-time-dep-k}
\end{eqnarray}

Here, we consider a universe labeled by $a$ with a uniformity and isotropy,
whose geometry and dynamics are described by the Robertson-Walker metric.
When the vacuum energy density $\rho_{\rm V}^{(a)} \equiv \mathscr{E}_{0}^{(a)}/V$
($V$:the volume of the universe) dominates over other energy densities,
the scale factor $a(t)$ varies based on the Friedmann equation:
\begin{eqnarray}
\left(\frac{\dot{a}(t)}{a(t)}\right)^2 = \frac{8\pi G_{\rm N}}{3} \rho_{\rm V}^{(a)},
\label{F-eq-k}
\end{eqnarray}
where $\dot{a}(t) \equiv {da(t)}/{dt}$
and $G_{\rm N}$ is gravitational constant.
Then, in the case with $\rho_{\rm V}^{(a)} > 0$,
an inflation (an exponential expansion of the universe)~\cite{Inf} occurs as
\begin{eqnarray}
a(t) = a(0) e^{H_{\rm V}^{(a)} t},
\label{inf-k}
\end{eqnarray}
where $H_{\rm V}^{(a)} \equiv \sqrt{{8\pi G_{\rm N}\rho_{\rm V}^{(a)}}/{3}}$.
In contrast, in the case with $\rho_{\rm V}^{(a)} < 0$, $a(t)$ oscillates.

In this way, we find that a universe can expand
at a very early stage and a macroscopic world can emerge
if a universe is dominated by a positive vacuum energy shortly after the birth.

\subsubsection{Third quantization}

In a universe labeled by $a$ whose spacetime varies, 
the spacetime itself is regarded as a dynamical object,
and a theory must be invariant under the general transformation of coordinates
and contain a gravitational field.

Let us study the invariance under the general transformation of coordinates
in our formulation.

Under the infinitesimal transformation
$x'^{\mu} = x^{\mu} - \varepsilon^{\mu}(x)$,
the field $\varphi_{k^{(a)}}$ transforms as 
$\varphi'_{k^{(a)}}(\bm{x}') = \varphi'_{k^{(a)}}(\bm{x}-\bm{\varepsilon}) 
= \varphi_{k^{(a)}}(\bm{x})$,
and then the change of $\varphi_{k^{(a)}}$ is given by
\begin{eqnarray}
\delta_{\varepsilon} \varphi_{k^{(a)}}(\bm{x}) 
= \varphi'_{k^{(a)}}(\bm{x}) - \varphi_{k^{(a)}}(\bm{x})
= \varphi_{k^{(a)}}(\bm{x}+\bm{\varepsilon}) - \varphi_{k^{(a)}}(\bm{x})
= \varepsilon^i(x) \partial_i \varphi_{k^{(a)}}(\bm{x}).
\label{delta-varphi-cl-k}
\end{eqnarray}
If the field's functional 
$\widehat{\varPhi}_{(a)}\left(\{\varphi_{k^{(a)}}\}, t\right)$ 
transforms as
\begin{eqnarray}
\widehat{\varPhi}'_{(a)}\left(\{\varphi'_{k^{(a)}}\}, t'\right) = 
\widehat{\varPhi}'_{(a)}\left(\{\varphi_{k^{(a)}} + \delta_{\varepsilon}\varphi_{k^{(a)}}\}, 
t-\varepsilon_t\right) 
= \widehat{\varPhi}_{(a)}\left(\{\varphi_{k^{(a)}}\}, t\right)
\label{varPhi(a)-trans}
\end{eqnarray}
under $x'^{\mu} = x^{\mu} - \varepsilon^{\mu}(x)$
with $\varepsilon^0(x) = c \varepsilon_t(x)$,
the change of $\widehat{\varPhi}_{(a)}\left(\{\varphi\}_{k^{(a)}}, t\right)$ 
is induced such that
\begin{align}
\delta_{\varepsilon} \widehat{\varPhi}_{(a)}\left(\{\varphi\}_{k^{(a)}}, t\right) 
&= \widehat{\varPhi}'_{(a)}\left(\{\varphi\}_{k^{(a)}}, t\right) 
- \widehat{\varPhi}_{(a)}\left(\{\varphi\}_{k^{(a)}}, t\right)
\nonumber \\
&= \widehat{\varPhi}'_{(a)}\left(\{\varphi_{k^{(a)}} - \delta_{\varepsilon}\varphi_{k^{(a)}}\}, 
t+\varepsilon_t\right)
- \widehat{\varPhi}_{(a)}\left(\{\varphi\}_{k^{(a)}}, t\right)
\nonumber \\
&= - \int d^3x \sum_{k^{(a)}}
\delta_{\varepsilon} \varphi_{k^{(a)}}(\bm{x}) \frac{\delta}{\delta \varphi_{k^{(a)}}(\bm{x})}
\widehat{\varPhi}_{(a)}\left(\{\varphi_{k^{(a)}}\}, t\right)
+ \varepsilon_t \frac{\partial}{\partial t}
\widehat{\varPhi}_{(a)}\left(\{\varphi_{k^{(a)}}\}, t\right)
\nonumber \\
&= - \int d^3x \sum_{k^{(a)}}
\varepsilon^i(x) \partial_i \varphi_{k^{(a)}} 
\frac{\delta}{\delta \varphi_{k^{(a)}}}
\widehat{\varPhi}_{(a)}\left(\{\varphi_{k^{(a)}}\}, t\right)
\nonumber \\
&~ ~~~~~~~~~~~
+ \varepsilon_t(x) \frac{\partial}{\partial t}
\widehat{\varPhi}_{(a)}\left(\{\varphi_{k^{(a)}}\}, t\right),
\label{delta-varPhi-k}
\end{align}
and then the change of the action integral is given by
\begin{align}
\delta_{\varepsilon} \widehat{S}_{\{\varPhi\}} 
&= \left[\widehat{\varPhi}^{\dagger}_{(a)}\left(\{\varphi_{k^{(a)}}\}, t\right)
\left(\frac{1}{c} \int d^3x~\varepsilon^{\mu}(x) {\widehat{T}^0}_{~\mu}(\bm{x})\right)
\widehat{\varPhi}_{(a)}\left(\{\varphi_{k^{(a)}}\}, t\right)\right]_{t_i}^{t_f}
\nonumber \\
&= \left[\widehat{\varPhi}^{\dagger}_{(a)}\left(\{\varphi_{k^{(a)}}\}\right)
\left(\frac{1}{c} \int d^3x~\varepsilon^{\mu}(x) {\widehat{T}^0}_{~\mu}(x)\right)
\widehat{\varPhi}_{(a)}\left(\{\varphi_{k^{(a)}}\}\right)\right]_{t_i}^{t_f},
\label{delta-SvarPhi-k}
\end{align}
using the Euler-Langrange equation.

When $\delta_{\varepsilon} \widehat{S}_{\{\varPhi\}} = 0$ holds
for arbitrary $\varepsilon^{\mu}(x)$, we have a physical state condition:
\begin{eqnarray}
\widehat{\varPhi}^{\dagger}_{(a)}\left(\{\varphi_{k^{(a)}}\}\right)
{\widehat{T}^0}_{~\mu}(x)\widehat{\varPhi}_{(a)}\left(\{\varphi_{k^{(a)}}\}\right)
|\varPsi_{\rm M_{II}}(t)\rangle = 0,
\label{phys-cond-k}
\end{eqnarray}
and it leads to the condition and the equation:
\begin{eqnarray}
{\widehat{T}^0}_{~\mu}(x)|\varPsi(t)\rangle_{\{\varphi_{k^{(a)}}\}} = 0
\label{phys-cond-a-k}
\end{eqnarray}
and 
\begin{eqnarray}
{\widehat{T}^0}_{~\mu}(x)\varPsi\left(\varphi_{k^{(a)}}, t\right) = 0,
\label{WW-eq-k}
\end{eqnarray}
respectively.
The time-component of eq.~\eqref{WW-eq-k} is equivalent to 
the Wheeler-deWitt equation~\cite{DW}
which represents the Hamiltonian constraint~\cite{ADM} at the quantum level, and 
we arrive at a fundamental formula on the third quantization~\cite{3rd}.

\subsubsection{Landscape}

First, we list postulates about a vacuum state and multiverse.
\begin{itemize}
\item In each universe, a vacuum state can change by a phase transition
(like the electroweak transition and QCD transition in our universe),
and hence there can appear a variety of vacuum states,
which are represented as $|0\rangle_{\{\varphi_{k^{(a)}}\}}$ together
for a universe with a set of elementary particles $\{\varphi_{k^{(a)}}\}$.
\item Even if a universe A produces a universe B~\cite{mini} using a mechanism such as
the eternal inflation~\cite{Einf}, 
as far as the mechanism works under physical laws in the universe A,
the universe B must obey the same physical laws of the universe A.
\end{itemize}

From the above postulates, we expect that a wider theoretical framework
is necessary to study a creation of universes with different
elementary particles and parameters and to discuss the relationship among them.
We investigate such a framework by reference to the landscape
of string theory vacua~\cite{Ls,Ls2}. 

Here, we list features of string theories and those vacua~\cite{GSW,P}.
\begin{itemize}
\item There are five kinds of superstring theories in ten-dimensional spacetime.
Numerous four-dimensional string models are constructed
after compactifying an extra six-dimensional space.
\item A variety of models originate from a diverse of structure of the extra space
and different configurations of strings.
Four-dimensional string models are specified by (the vacuum expectation values of)
scalar fields $\varphi_{{\rm L}m}$ including moduli
which characterize the structure of the extra space.
The space spanned by $\varphi_{{\rm L}m}$ is called ``landscape'',
and its altitude corresponds to a value of vacuum energy.
The landscape can be regarded as a potential.
\item Particle contents in each string model
are stored in string fields symbolically
denoted by $\widehat{\varphi}_{\rm st}(\bm{X}(\sigma), \bm{\psi}(\sigma))$.
\end{itemize}

Let us suppose that there is a Hamiltonian $\widehat{H}_{\{\varphi_{\rm L}\}}$
containing a potential $\widehat{V}_{\{\varphi_{\rm L}\}}$ whose local minimums
correspond to solutions describing universes with different particle contents
and physical parameters, and the dynamics can be compactly expressed by the action integral:
\begin{align}
\widehat{S}_{\varPhi_{\rm L}}
&= \int \widehat{L}_{\varPhi_{\rm L}}dt
\nonumber \\
&= \int \left(\widehat{\varPhi}_{\rm L}^{\dagger}\left(\{\varphi_{\rm L}\}, t\right) 
i\hbar \frac{\partial}{\partial t}\widehat{\varPhi}_{\rm L}\left(\{\varphi_{\rm L}\}, t\right)
- \widehat{\varPhi}_{\rm L}^{\dagger}\left(\{\varphi_{\rm L}\}, t\right) 
\widehat{H}_{\{\varphi_{\rm L}\}}
\widehat{\varPhi}_{\rm L}\left(\{\varphi_{\rm L}\}, t\right)\right)dt,
\label{S-varPhi-L}
\end{align}
where $\{\varphi_{\rm L}\}$ is a set of fields $\varphi_{{\rm L}m}$ 
living on the landscape,
and $\widehat{\varPhi}_{\rm L}\left(\{\varphi_{\rm L}\}, t\right)$
is the functional operator concerning $\{\varphi_{\rm L}\}$.

The functional operator satisfies the equation:
\begin{eqnarray}
i\hbar \frac{\partial}{\partial t}\widehat{\varPhi}_{\rm L}\left(\{\varphi_{\rm L}\}, t\right) 
= \widehat{H}_{\{\varphi_{\rm L}\}} \widehat{\varPhi}_{\rm L}\left(\{\varphi_{\rm L}\}, t\right),
\label{H-eq-varPhi-L}
\end{eqnarray}
and, from the stationary condition:
\begin{eqnarray}
\frac{\delta}{\delta \varphi_{{\rm L}m}}\widehat{H}_{\{\varphi_{\rm L}\}} = 0,
\label{stational-varPhi-L}
\end{eqnarray}
we obtain solutions $\{\varphi_{\rm L}^{(a)}\}$ which are identified with
models owning a set of elementary particles $\{\varphi_{k^{(a)}}\}$.
Then, $\widehat{\varPhi}_{\rm L}\left(\{\varphi_{\rm L}^{(a)}\}, t\right)$ becomes 
$\widehat{\varPhi}_{(a)}\left(\{\varphi_{k^{(a)}}\}, t\right)$,
and, putting together them,
we arrive at an effective formulation described by the action integral \eqref{S-varPhi-k}.

As a mechanism to produce the level II multiverse,
the eternal inflation using the potential $\widehat{V}_{\{\varphi_{\rm L}\}}$
in the landscape looks promising.
Then, some of $\varphi_{{\rm L}m}$ can play a role of inflatons.
For more detail, a vacuum state $|0\rangle_{\{\varphi_{\rm L}\}}$ and a spacetime
can emerge by the operation of 
$\widehat{\varPhi}^{\dagger}_{\rm L}\left(\{\varphi_{\rm L}\}, t\right)$
such as $|0\rangle_{\{\varphi_{\rm L}\}}=
\widehat{\varPhi}^{\dagger}_{\rm L}\left(\{\varphi_{\rm L}\}, t\right)|\Slash\rangle$.
Each region of the space expands while $\varphi_{{\rm L}m}$ are rolling down towards
a different local minimum of the potential
or they are transitioning from a local minimum to another one.
When the potential has flat directions for some $\varphi_{{\rm L}m}$,
those values can correspond to values of parameters.
After $\varphi_{{\rm L}m}$ arrive at the local minimum labeled by $\{\varphi_{\rm L}^{(a)}\}$
including a set of parameters,
the values of $\varphi_{{\rm L}m}$ are fixed,
the vacuum $|0\rangle_{\{\varphi_{\rm L}\}}$ changes into $|0\rangle_{\{\varphi_{k^{(a)}}\}}$,
and a universe with $|0\rangle_{\{\varphi_{k^{(a)}}\}}$ comes into existence.
If the universe is dominated by a positive vacuum energy 
coming from some particle (called an inflaton) in $\{\varphi_{k^{(a)}}\}$,
an inflation can occurs in the universe.
After the inflation comes to an end, 
the inflaton decays into other particles in $\{\varphi_{k^{(a)}}\}$
and the universe becomes extremely hot and dense (like a big bang in our universe).
In this way, various universes with different particle contents and parameters
(different physical laws) can be generated.

A change between universes with different physical laws 
can also occur and its rate can be estimated
in the setup based on the action integral \eqref{S-varPhi-L},
if there are no selection rules to forbid the change.
According to the laws of quantum physics,
we can calculate the transition amplitude $\widehat{S}_{\rm f~\!i}$
from a vacuum state in a universe
with $\{\varphi_{k^{(a)}}\}$ to a different one with $\{\varphi_{k^{(b)}}\}$
by using the formula:
\begin{align}
\widehat{S}_{\rm f~\!i} =
\langle\Slash |\widehat{\varPhi}_{(b)}\left(\{\varphi_{k^{(b)}}\}, t_{\rm f}\right)
~e^{-\frac{i}{\hbar}\widehat{H}_{\varPhi_{\rm L}}(t_{\rm f}-t_{\rm i})}~
\widehat{\varPhi}^{\dagger}_{(a)}\left(\{\varphi_{k^{(a)}}\}, t_{\rm i}\right)|\Slash\rangle,
\label{(a)to(b)}
\end{align}
where $\widehat{H}_{\varPhi_{\rm L}}
= \widehat{\varPhi}_{\rm L}^{\dagger}\left(\{\varphi_{\rm L}\}, t\right) 
\widehat{H}_{\{\varphi_{\rm L}\}}
\widehat{\varPhi}_{\rm L}\left(\{\varphi_{\rm L}\}, t\right)$.

\section{Conclusions and discussions}

Taking the embedded structure in QFT as a guiding principle,
we have arrived at a wider theoretical framework called
quantum field's functional (QFFT) that has QFT built-in.
QFFT is constructed by using functional operators
and has the nested structure\footnote{
Another extension of QFT owning a nested structure
has been carried out by promoting a wave functional in QFT
into an operator in reference to the second quantization \cite{MP}.
In this formulation, the operator can create or annihilate a multi-particle state.}.

A rationale behind such a structure is 
that a physical state should be evolved by the same type of equation:
$i\hbar d|\varPsi_{\bigstar}(t)\rangle/dt 
= \widehat{H}_{\star}|\varPsi_{\bigstar}(t)\rangle$
for any quantum systems in a world where the laws of quantum theory hold.
In concrete, for a system in the level II multiverse,
it is given by $i\hbar d|\varPsi_{\rm M_{II}}(t)\rangle/dt 
= \widehat{H}_{\{\varPhi\}}|\varPsi_{\rm M_{II}}(t)\rangle$.
For a system that the number of particles can vary,
it becomes $i\hbar d|\varPsi(t)\rangle/dt 
= \widehat{H}_{\varphi}|\varPsi(t)\rangle$ based on QFT.
For a system with a definite number of particles,
it takes $i\hbar d|\psi(t)\rangle/dt 
= \widehat{H}|\psi(t)\rangle$ based on QM.
Hence the Schr\"{o}dinger equation for a state is worshiped 
as a master equation, and it makes us think 
that nature is not all that complicated at a fundamental level.

On the one hand, QFFT would not be well-defined mathematically 
in the same way as QFT, that is, it is not yet known 
whether functional operators can be rigorously defined or not.
On the other hand, QFFT has several advantages.
First, it is relatively naturally understood 
that fermions (fields as Grassmann variables)
and bosons are represented by field operators
satisfying the anti-commutation relations and the commutation relations, respectively.
Second, the which-came-first-particles-or-spacetime problem 
(which-came-first-physical-laws-or-our-universe problem) can be solved,
if particles and space-time (physical laws and our universe) 
can be installed as a unit by a functional operator.
Third, the level II multiverse could be described by QFFT.
Last, topics related to a beginning of the universe 
such as an inflation, the third quantization
and the landscape can be studied in our formulation.
Our framework might also be applied to the level III multiverse
where the level III multiverse means quantum mechanical many worlds~\cite{EIII}\footnote{
It is proposed that the eternally inflating multiverse
and quantum mechanical many worlds are the same thing~\cite{N}.}.

At present, QFFT is merely an effective framework to formulate laws of particle physics
or a tool in understanding of quantum physics in the same way as QFT, 
and hence it might possess not so powerful predictability. 
It is intriguing to examine whether QFFT has both predictability and falsifiability or not,
based on investigations of the level II multiverse.

There is a possibility that QFFT plays a supporting role as follows. 
For several decades, the exploration of a fundamental theory
has been actively carried out by using two approaches
such as the bottom-up one (which takes a route 
from the standard model to a theory beyond the standard model)
and the top-down one (which takes a route from a theory of everything to the standard model).
Superstring theory has been a hopeful candidate of an ultimate theory,
and it offers vast numbers of string models 
which can be identified with members of the level II multiverse.
Even if current superstring theory did not answer the question ``How are universes created?'',
it would be open to extend its framework.
Then, it is expected that QFFT serves as a good model.
Furthermore, a combination of above-mentioned two approaches
and an extension of framework would be promising to unravel secrets on our universe
and beyond it.

\section*{Acknowledgments}
This work was supported in part by scientific grants 
from the Ministry of Education, Culture,
Sports, Science and Technology under Grant No.~22K03632.

\appendix

\section{Hamiltonian operators in QM, QFT and QFFT}

We give explicit forms of Hamiltonian operators in QM, QFT and QFFT
for several cases.

\subsection{Hamiltonian operators in non-relativistic quantum theory}

For a non-relativistic electron with a mass $m_{\rm e}$,
the Hamiltonian operator in QM is given by
\begin{eqnarray}
\widehat{H}(\bm{x},-i\hbar\bm{\nabla}) = - \frac{\hbar^2}{2m_{\rm e}} \bm{\nabla}^2 + V(\bm{x})
\label{A-H-NR-QM}
\end{eqnarray}
and then the Hamiltonian operators in QFT and QFFT are given by
\begin{eqnarray}
\widehat{H}_{\varphi} 
= \int \widehat{\varphi}^{\dagger}(x)\widehat{H}\widehat{\varphi}(x) d^3x
= \int \widehat{\varphi}^{\dagger}(x) \left(- \frac{\hbar^2}{2m_{\rm e}} \bm{\nabla}^2
+ V(\bm{x})\right)
\widehat{\varphi}(x) d^3x
\label{A-H-NR-QFT}
\end{eqnarray}
and
\begin{eqnarray}
\widehat{H}_{\varPhi}
= \widehat{\varPhi}^{\dagger}(\{\varphi\},t) 
\left\{\int \frac{\delta}{\delta \varphi(\bm{x})}
 \left(-\frac{\hbar^2}{2m_{\rm e}} \bm{\nabla}^2 + V(\bm{x})\right)
{\varphi}(\bm{x}) d^3x\right\}
\widehat{\varPhi}(\{\varphi\},t),
\label{A-H-NR-QFFT}
\end{eqnarray}
respectively.

\subsection{Hamiltonian operators in relativistic quantum theory}

(1) Weyl fermion\\
For a free left-handed Weyl fermion (a massless particle with a helicity $-1/2$), 
the Hamiltonian operator in QM is given by
\begin{eqnarray}
\widehat{H}(\bm{x}, i\hbar\bm{\nabla}) 
= - i \hbar c \bm{\sigma} \cdot \bm{\nabla}
\label{A-H-Weyl-QM}
\end{eqnarray}
and then that in QFT is given by
\begin{eqnarray}
\widehat{H}_{\varphi} 
= \int \widehat{\varphi}^{\dagger}(x)\widehat{H}\widehat{\varphi}(x) d^3x
= \int \widehat{\varphi}^{\dagger}(x) \left(- i \hbar c \bm{\sigma} \cdot \bm{\nabla}\right)
\widehat{\varphi}(x) d^3x,
\label{A-H-Weyl-QFT}
\end{eqnarray}
where $\bm{\sigma}$ is Pauli matrices.
For a free right-handed Weyl fermion (a massless particle with a helicity $1/2$), 
$\widehat{H}(\bm{x}, i\hbar\bm{\nabla}) 
= i \hbar c \bm{\sigma} \cdot \bm{\nabla}$ and
$\displaystyle{\widehat{H}_{\varphi} 
= \int \widehat{\varphi}^{\dagger}(x) \left(i \hbar c \bm{\sigma} \cdot \bm{\nabla}\right)
\widehat{\varphi}(x) d^3x}$
are obtained by replacing $\bm{\sigma}$ into $-\bm{\sigma}$ in eqs.~\eqref{A-H-Weyl-QM}
and \eqref{A-H-Weyl-QFT}.

The Hamiltonian operator in QFFT is given by
\begin{eqnarray}
\widehat{H}_{\varPhi}
= \widehat{\varPhi}^{\dagger}(\{\varphi\})  
\left\{\int \frac{\delta}{\delta \varphi(\bm{x})} 
\left(- i \hbar c \bm{\sigma} \cdot \bm{\nabla}\right)
{\varphi}(\bm{x}) d^3x\right\}
\widehat{\varPhi}(\{\varphi\})
\label{A-H-Weyl-QFFT}
\end{eqnarray}
for a left-handed Weyl fermion.

~~\\
(2) Dirac fermion\\
For a free Dirac fermion (a particle with a mass $m$ and spin 1/2), 
the Hamiltonian in QM is given by
\begin{eqnarray}
\widehat{H}(\bm{x}, i\hbar\bm{\nabla}) 
= - i \hbar c \bm{\alpha} \cdot \bm{\nabla}+ \beta mc^2,
\label{A-H-Dirac-QM}
\end{eqnarray}
and then the Hamiltonian operators in QFT and QFFT are given by
\begin{eqnarray}
\widehat{H}_{\varphi} 
= \int \widehat{\varphi}^{\dagger}(x)\widehat{H}\widehat{\varphi}(x) d^3x
= \int \widehat{\varphi}^{\dagger}(x) \left(- i \hbar c \bm{\alpha} \cdot \bm{\nabla}
+ \beta mc^2\right)
\widehat{\varphi}(x) d^3x
\label{A-H-Dirac-QFT}
\end{eqnarray}
and 
\begin{eqnarray}
\widehat{H}_{\varPhi}
= \widehat{\varPhi}^{\dagger}(\{\varphi\})  
\left\{\int \frac{\delta}{\delta \varphi(\bm{x})} 
\left(- i \hbar c \bm{\alpha} \cdot \bm{\nabla} + \beta mc^2\right)
{\varphi}(\bm{x}) d^3x\right\}
\widehat{\varPhi}(\{\varphi\}),
\label{A-H-Dirac-QFFT}
\end{eqnarray}
respectively.

\section{Wave functional}

A wave functional in QFT is defined by
\begin{eqnarray}
\varPsi(\varphi, t) \equiv \langle \varphi|\varPsi(t)\rangle,
\label{B-varPsi}
\end{eqnarray}
where $\langle \varphi|$ is a bra-vector and its ket-vector $|\varphi\rangle$
is an eigenvector satisfying
\begin{eqnarray}
\widehat{\varphi}(\bm{x})|\varphi\rangle=\varphi(\bm{x})|\varphi\rangle,~~
\langle\varphi |\widehat{\varphi}^{\dagger}(\bm{x})
=\langle\varphi |{\varphi}^{\dagger}(\bm{x})
\label{B-|varpsi>}
\end{eqnarray}
and $\varphi(\bm{x})$ is a configuration of classical field.
For fermion, $\varphi(\bm{x})$ takes Grassmann values.
The eigenvectors $\langle \varphi|$ and $|\widetilde{\varphi}\rangle$ 
satisfy the relations:
\begin{eqnarray}
\langle \varphi|\widetilde{\varphi}\rangle = \delta(\varphi - \widetilde{\varphi})
= \int \mathscr{D}\alpha~
e^{i\int \alpha(\bm{x})(\varphi(\bm{x})-\widetilde{\varphi}(\bm{x}))d^3x},~~
\int \mathscr{D}\varphi~|\varphi\rangle\langle \varphi| = 1.
\label{B-|varpsi>-rels}
\end{eqnarray}

The expectation value of $\widehat{H}_{\varphi}$ is written by
\begin{align}
& \langle\varPsi(t)|\widehat{H}_{\varphi}|\varPsi(t)\rangle
= \int \mathscr{D}\varphi~ \int \mathscr{D}\widetilde{\varphi}~
\langle\varPsi(t)|\varphi\rangle\langle \varphi|\widehat{H}_{\varphi} 
|\widetilde{\varphi}\rangle\langle\widetilde{\varphi}|\varPsi(t)\rangle
\nonumber \\
&~~~~= \int \mathscr{D}\varphi~ \int \mathscr{D}\widetilde{\varphi}~
\varPsi^{\dagger}(\varphi, t) 
\widehat{H}_{\varphi}\delta(\varphi - \widetilde{\varphi})
\varPsi(\widetilde{\varphi}, t)
= \int \mathscr{D}\varphi~
\varPsi^{\dagger}(\varphi, t)\widehat{H}_{\varphi}\varPsi(\varphi, t),
\label{B-<H>}
\end{align}
using eq.~\eqref{B-|varpsi>-rels}.

A general solution  of the wave functional is given by
\begin{eqnarray}
\varPsi(\varphi, t) = \sum_{N=0}^{\infty}
e^{-\frac{i}{\hbar}\mathscr{E}_N t} \mathcal{U}_N(\varphi),
\label{B-varPsi-ex}
\end{eqnarray}
where $\mathscr{E}_N$ and $\mathscr{U}_N(\varphi)$
are eigenvalues and eigenvectors of the eigenvalue equation:
\begin{eqnarray}
\widehat{H}_{\varphi}\mathcal{U}_N(\varphi)
= \mathscr{E}_N\mathcal{U}_N(\varphi).
\label{B-H-eigen}
\end{eqnarray}
Note that $\varPsi(\varphi, t)$ satisfies eq.~\eqref{S-eq-QFT}, i.e., 
$i \hbar {\partial\varPsi(\varphi, t)}/{\partial t} 
= \widehat{H}_{\varphi} \varPsi(\varphi, t)$

In the $\bm{x}$-representation, the state $|\varPsi(t)\rangle$ is expressed by
\begin{eqnarray}
|\varPsi(t)\rangle = \psi^{(0)}(t)|0\rangle 
+ \sum_{N = 1}^{\infty} \int d^3x_1 \cdots d^3x_N
\psi(\bm{x}_1, \cdots, \bm{x}_N, t) |\bm{x}_1, \cdots, \bm{x}_N\rangle,
\label{B-state}
\end{eqnarray}
where $|0\rangle$ is a vacuum state, 
$\psi(\bm{x}_1, \cdots, \bm{x}_N, t)$ is a wave function
of $N$-particle states and $|\bm{x}_1, \cdots, \bm{x}_N\rangle$
is a ket vector which satisfies
$\widehat{\bm{x}}_l |\bm{x}_1, \cdots, \bm{x}_N\rangle
= \bm{x}_l|\bm{x}_1, \cdots, \bm{x}_N\rangle$ 
($l=1,\cdots, N$) and is defined by
\begin{eqnarray}
|\bm{x}_1, \cdots, \bm{x}_N\rangle
\equiv \frac{1}{\sqrt{N!}} \widehat{\varphi}^{\dagger}(\bm{x}_1) 
\cdots \widehat{\varphi}^{\dagger}(\bm{x}_N) 
|0\rangle,
\label{B-x-state}
\end{eqnarray}
using $\widehat{\varphi}^{\dagger}(\bm{x})$
which plays a role of a creation operator of a particle $\varphi$ 
or an annihilation operator of its anti-particle $\overline{\varphi}$.

Multiplying $\langle \varphi|$ by both side of eq.~\eqref{B-state}
and using eq.~\eqref{B-|varpsi>},
the following relation is obtained,
\begin{align}
\varPsi(\varphi, t) &=
\langle \varphi|\varPsi(t)\rangle 
\nonumber \\
&= \psi^{(0)}(t)\langle \varphi|0\rangle 
+ \sum_{N = 1}^{\infty} \int d^3x_1 \cdots d^3x_N
\psi(\bm{x}_1, \cdots, \bm{x}_N, t)\langle \varphi|\bm{x}_1, \cdots, \bm{x}_N\rangle
\nonumber \\
&= \psi^{(0)}(t)\langle \varphi|0\rangle 
\nonumber \\
&~ ~~ + \sum_{N = 1}^{\infty} \int d^3x_1 \cdots d^3x_N
\psi(\bm{x}_1, \cdots, \bm{x}_N, t) \frac{1}{\sqrt{N!}}
\varphi^{\dagger}(\bm{x}_1) \cdots \varphi^{\dagger}(\bm{x}_N) \langle\varphi|0\rangle.
\label{B-state-<varphi|}
\end{align} 

From eqs.~\eqref{B-varPsi-ex} and \eqref{B-state-<varphi|},
by taking $\mathcal{V}_0(\varphi) = \langle \varphi|0\rangle$
and $\mathcal{U}_N(\varphi) \equiv C_N(\varphi) \mathcal{V}_0(\varphi)$,
we obtain the relations:
\begin{eqnarray}
&~& e^{-\frac{i}{\hbar}\mathscr{E}_0 t}C_{0}(\varphi) = \psi^{(0)}(t),~~
\label{B-U0}\\
&~& e^{-\frac{i}{\hbar}\mathscr{E}_N t}C_{N}(\varphi)
= \int d^3x_1 \cdots d^3x_N
\psi(\bm{x}_1, \cdots, \bm{x}_N, t) \frac{1}{\sqrt{N!}}
\varphi^{\dagger}(\bm{x}_1) \cdots \varphi^{\dagger}(\bm{x}_N).
\label{B-UN}
\end{eqnarray}

Multiplying $|\varphi\rangle$ by $\langle \bm{x}|$, we obtain the relation:
\begin{eqnarray}
\langle \bm{x}|\varphi\rangle = \langle 0|\widehat{\varphi}(\bm{x})|\varphi\rangle
=\varphi(\bm{x})\langle 0|\varphi\rangle 
= \varphi(\bm{x})\mathcal{V}_0^{\dagger}(\varphi)
\label{B-<x|varphi>}
\end{eqnarray}
with $\mathcal{V}_0^{\dagger}(\varphi)=\langle 0|\varphi\rangle$.

Here, we present formula including 
$\mathcal{V}_0(\varphi)$ and $\mathcal{V}_0^{\dagger}(\varphi)$,
\begin{align}
&~ \int \mathscr{D}\varphi~
\mathcal{V}_0^{\dagger}(\varphi)\mathcal{V}_0(\varphi)
= \int \mathscr{D}\varphi~\langle 0|\varphi\rangle\langle\varphi|0\rangle 
= \langle 0|0\rangle =1,~~
\label{B-formula1}\\
&~ \int \mathscr{D}\varphi~\frac{1}{N!}\varphi(\bm{x}_1)\cdots\varphi(\bm{x}_N)
\varphi^{\dagger}(\bm{x}'_1)\cdots\varphi^{\dagger}(\bm{x}'_N)
\mathcal{V}_0^{\dagger}(\varphi)\mathcal{V}_0(\varphi)
\nonumber \\
&~ ~~~~~~~~~~~~~ = \delta^3(\bm{x}_1-\bm{x}'_1)\cdots\delta^3(\bm{x}_N-\bm{x}'_N),~~
\label{B-formula2}\\
&~ \mathcal{V}_0(\varphi)\mathcal{V}^{\dagger}_0(\widetilde{\varphi})
+ \sum_{N = 1}^{\infty} \int d^3x_1 \cdots d^3x_N~\frac{1}{N!}
\varphi^{\dagger}(\bm{x}_1) \cdots \varphi^{\dagger}(\bm{x}_N)
\widetilde{\varphi}(\bm{x}_1) \cdots \widetilde{\varphi}(\bm{x}_N)
\mathcal{V}_0(\varphi)\mathcal{V}^{\dagger}_0(\widetilde{\varphi})
\nonumber \\
&~  ~~~~~~~~~~~~~ = \delta(\varphi-\widetilde{\varphi}).
\label{B-formula3}
\end{align}

As a reference, 
the vacuum wave function
$\varPsi_{\rm V}(t)=\langle 0|\varPsi(t)\rangle
= \psi^{(0)}(t)$ is written by
\begin{eqnarray}
\varPsi_{\rm V}(t)=\psi^{(0)}(t)=\langle 0|\varPsi(t)\rangle
= \int \mathscr{D}\varphi~ \langle 0|\varphi\rangle
\langle\varphi|\varPsi(t)\rangle
= \int \mathscr{D}\varphi~ \mathcal{V}_0^{\dagger}(\varphi)\varPsi(\varphi, t).
\label{B-varPsiV}
\end{eqnarray}

In the same way, a component of functional
$\langle \bm{x}_1, \cdots, \bm{x}_N|\varPsi(t)\rangle 
= \psi(\bm{x}_1, \cdots, \bm{x}_N, t)$ is written by
\begin{eqnarray}
\psi(\bm{x}_1, \cdots, \bm{x}_N, t)
= \int \mathscr{D}\varphi~\frac{1}{\sqrt{N!}}\varphi(\bm{x}_1) \cdots \varphi(\bm{x}_N)~
\mathcal{V}_0^{\dagger}(\varphi)\varPsi(\varphi, t).
\label{B-varPsi-psi}
\end{eqnarray}

The state $|\varPsi(t)\rangle$ is rewritten as
\begin{align}
|\varPsi(t)\rangle 
&= \psi^{(0)}(t) |0\rangle
+ \sum_{N = 1}^{\infty} \int d^3x_1 \cdots d^3x_N
\psi(\bm{x}_1, \cdots, \bm{x}_N, t) \frac{1}{\sqrt{N!}}
\widehat{\varphi}^{\dagger}(\bm{x}_1)
 \cdots \widehat{\varphi}^{\dagger}(\bm{x}_N)|0\rangle
\nonumber \\
&= \psi^{(0)}(t) |0\rangle
\nonumber \\
&~  + \sum_{N = 1}^{\infty} \int d^3x_1 \cdots d^3x_N
\psi(\bm{x}_1, \cdots, \bm{x}_N, t) \frac{1}{\sqrt{N!}}
\int \mathscr{D}\varphi|\varphi\rangle\langle\varphi|
\widehat{\varphi}^{\dagger}(\bm{x}_1)
 \cdots \widehat{\varphi}^{\dagger}(\bm{x}_N)|0\rangle
\nonumber \\
&= \psi^{(0)}(t) |0\rangle
+ \sum_{N = 1}^{\infty} \int d^3x_1 \cdots d^3x_N
\psi(\bm{x}_1, \cdots, \bm{x}_N, t) \frac{1}{\sqrt{N!}}
\varphi^{\dagger}(\bm{x}_1)
 \cdots \varphi^{\dagger}(\bm{x}_N)|0\rangle
\nonumber \\
&= \varPsi(\{\varphi\}, t)|0\rangle,
\label{B-state-rewritten}
\end{align}
where ${\varPsi}(\{\varphi\},t)$ is a wave functional defined by
\begin{eqnarray}
{\varPsi}(\{\varphi\},t)
\equiv \psi^{(0)}(t)+
\sum_{N = 1}^{\infty} \int d^3x_1 \cdots d^3x_N
\psi(\bm{x}_1, \cdots, \bm{x}_N, t) \frac{1}{\sqrt{N!}}
\varphi^{\dagger}(\bm{x}_1) \cdots \varphi^{\dagger}(\bm{x}_N).
\label{B-fuctional-M-1}
\end{eqnarray}
From eqs.~\eqref{B-state-<varphi|} and \eqref{B-fuctional-M-1},
we obtain the relation:
\begin{eqnarray}
\varPsi(\varphi, t) = {\varPsi}(\{\varphi\},t) \mathscr{V}_0(\varphi).
\label{B-fuctional-M-1-QFT}
\end{eqnarray}

For an operator $\widehat{\varOmega}_{\varphi}$,
it is shown that the following formula holds,
\begin{align}
\langle\varPsi(t)|\widehat{\varOmega}_{\varphi}|\varPsi(t)\rangle
&= \int \mathscr{D}\varphi~
\varPsi^{\dagger}(\varphi, t)\widehat{\varOmega}_{\varphi}\varPsi(\varphi, t)
\nonumber \\
&= \varPsi^{\dagger}(\{\varphi\}, t)\widehat{\varOmega}_{\varphi}\varPsi(\{\varphi\}, t),
\label{B-<Omega>}
\end{align}
using eq.~\eqref{B-formula1}.

\end{document}